%% file: Main.tex
\documentclass[10pt,twocolumn,letterpaper]{article}
\PassOptionsToPackage{dvipsnames,table}{xcolor}

\usepackage{cvpr}
\input{preamble}

\definecolor{cvprblue}{rgb}{0.21,0.49,0.74}

\usepackage{times}
\usepackage{graphicx}
\usepackage{amsmath}
\usepackage{amssymb}
\usepackage{booktabs}
\usepackage{rotating}
\usepackage{paralist}
\usepackage{caption}
\usepackage{subcaption}
\usepackage[linesnumbered,boxed]{algorithm2e}
\usepackage{arydshln}
\usepackage{array}
\usepackage{multirow}
\usepackage{ragged2e}
\usepackage{booktabs}
\usepackage{arydshln}
\usepackage{makecell}

\usepackage[dvipsnames]{xcolor}
\usepackage[accsupp]{axessibility}
\usepackage{times}
\usepackage{array}
\usepackage{ragged2e}
\usepackage{booktabs}
\usepackage{wrapfig}

\usepackage{graphicx}
\usepackage{amsmath}
\usepackage{amssymb}
\usepackage[pagebackref=true,breaklinks=true,colorlinks,bookmarks=false,citecolor=blue,linkcolor=blue]{hyperref}
\usepackage{makecell}
\usepackage[accsupp]{axessibility}
\usepackage{color, colortbl}
\definecolor{LightCyan}{rgb}{0.88,1,1}

\hypersetup{
    colorlinks=true,
    urlcolor=magenta,
    citecolor=blue,
}

\usepackage[T1]{fontenc}
\usepackage{multirow}
\usepackage{booktabs}
\usepackage{comment}
\usepackage{color}
\usepackage{amsmath}

\usepackage{textcomp}
\usepackage{siunitx}
\usepackage{tabulary}
\usepackage{makecell}
\usepackage{multirow}
\usepackage{booktabs}
\usepackage{hyperref}

\makeatletter
\newcommand\figref{Figure~\ref}
\newcommand{\tabref}[1]{Table~\ref{#1}}
\newcolumntype{P}[1]{>{\centering\arraybackslash}p{#1}}
\newcolumntype{M}[1]{>{\centering\arraybackslash}m{#1}}
\let\ts@includegraphics\includegraphics

\usepackage{float}
\usepackage{lipsum}   
\usepackage{rotating} 
\usepackage{stfloats} 
\usepackage[export]{adjustbox} 


\usepackage[capitalize]{cleveref}
\crefname{section}{Sec.}{Secs.}
\Crefname{section}{Section}{Sections}
\Crefname{table}{Table}{Tables}
\crefname{table}{Tab.}{Tabs.}

\def\paperID{9427} 
\def\confName{CVPR}
\def\confYear{2025}

\begin{document}

\title{\method{}: Unified Perceptual and Task-Oriented Image Restoration Model Using Diffusion Prior}



\vspace{-20mm}\author{
    I-Hsiang Chen\textsuperscript{1*}
    \quad Wei-Ting Chen\textsuperscript{1,2,4*}
    \quad Yu-Wei Liu\textsuperscript{1} 
    \quad Yuan-Chun Chiang\textsuperscript{1} \\
    \quad Sy-Yen Kuo\textsuperscript{1,3}
    \quad Ming-Hsuan Yang\textsuperscript{4,5}
    \\\\
    \hspace{-8mm}\textsuperscript{1}National Taiwan University\quad \textsuperscript{2}Microsoft\quad \textsuperscript{3}Chang Gung University\quad \\ \textsuperscript{4}UC Merced\quad \textsuperscript{5}Google Research
}

\maketitle


\begin{abstract}
Image restoration aims to recover content from inputs 
degraded by various factors, such as adverse weather, blur, and noise. Perceptual Image Restoration (PIR) methods improve visual quality but often do not support downstream tasks effectively. On the other hand, Task-oriented Image Restoration (TIR) methods focus on enhancing image utility for high-level vision tasks, sometimes compromising visual quality. This paper introduces \method{}, a unified image restoration model that bridges the gap between PIR and TIR by using a diffusion prior. The diffusion prior is designed to generate images that align with human visual quality preferences, but these images are often unsuitable for TIR scenarios. To solve this limitation, \method{} utilizes encoder features from an autoencoder to adapt the diffusion prior to specific tasks. We propose a Complementary Feature Restoration Module (CFRM) to reconstruct degraded encoder features and a Task Feature Adapter (TFA) module to facilitate adaptive feature fusion in the decoder. This design allows \method{} to optimize images for both human perception and downstream task requirements, addressing discrepancies between visual quality and functional needs. Integrating these modules also enhances \method{}'s adaptability and efficiency across diverse tasks. Extensive experiments demonstrate the superior performance of \method{} in both PIR and TIR scenarios.

\end{abstract}

\newcommand\blfootnote[1]{%
\begingroup
\renewcommand\thefootnote{}\footnote{#1}%
\addtocounter{footnote}{-1}%
\endgroup
}

\blfootnote{Project page: \url{https://unirestore.github.io/}}
\blfootnote{* Indicates equal contribution.}

\setlength{\parskip}{0em}

\input{Introduction}

\input{Related_Work}

\input{Methodology}

\input{Implementation_Detail}

\input{Experimental_Results}

\input{Conclusion}
\clearpage

{\small
\bibliographystyle{ieee_fullname}
\bibliography{egbib}
}

\end{document}

%% file: preamble.tex
\usepackage[dvipsnames]{xcolor}

\newcommand{\method}[1]{UniRestore}


%% file: Introduction.tex
\section{Introduction}
Image restoration~\cite{valanarasu2022transweather, potlapalli2024promptir} aims to restore content degraded by various factors, such as adverse weather, blur, and noise. These factors often reduce image perceptual visibility~\cite{wang2024data, li2020all} and negatively impact the performance of high-level vision applications, such as object detection~\cite{zhang2023adaptive,xu2021exploring} and semantic segmentation~\cite{saravanarajan2023improving,qian2024allweathernet}. Various restoration methods have been developed and studied over the past decades to address this wide range of image restoration challenges. These include methods focused on improving perceptual quality~\cite{ren2023multiscale,zamir2021multi} as well as those designed to enhance the performance of downstream tasks~\cite{son2020urie,yang2023visual}.

\begin{figure}[t!]
    \centering
    \vspace{-5pt}
    \begin{subfigure}{0.4\columnwidth}
        \centering
        \includegraphics[width=\linewidth]{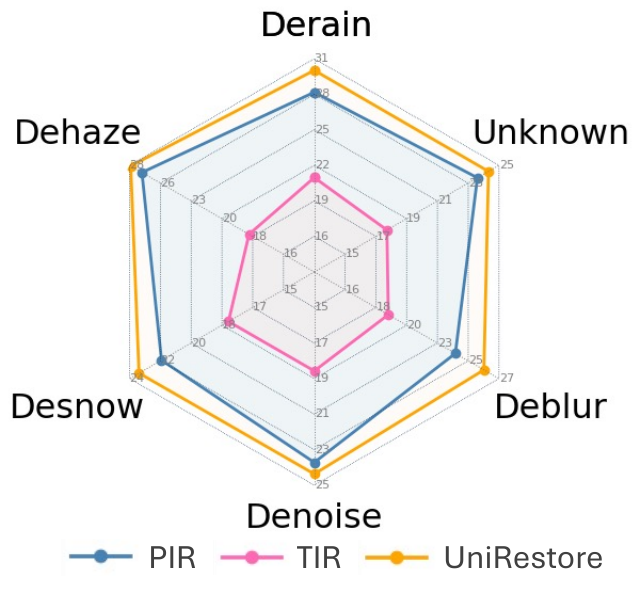}
        \caption{Perceptual IR}
        \label{fig:teaser_pir}
    \end{subfigure}
    \hspace{0.1cm} 
    \begin{subfigure}{0.5\columnwidth}
        \centering
        \includegraphics[width=\linewidth]{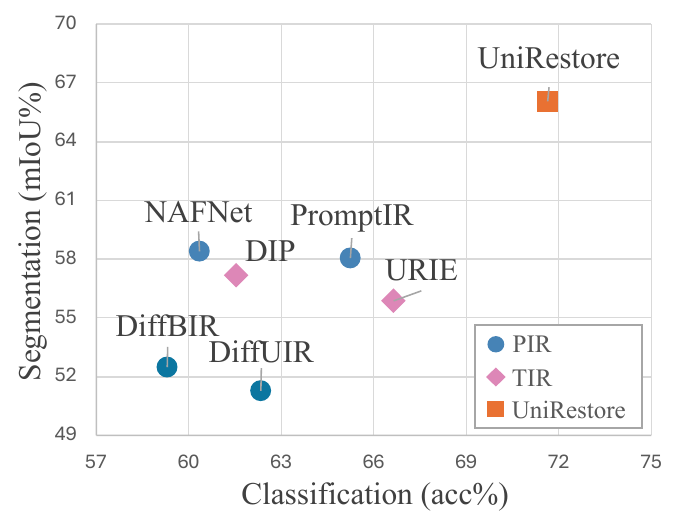}
        \caption{Task-oriented IR}
        \label{fig:teaser_TIR}
    \end{subfigure}
    \vspace{-5pt}
    \begin{subfigure}{\columnwidth}
        \centering
        \includegraphics[width=\linewidth]{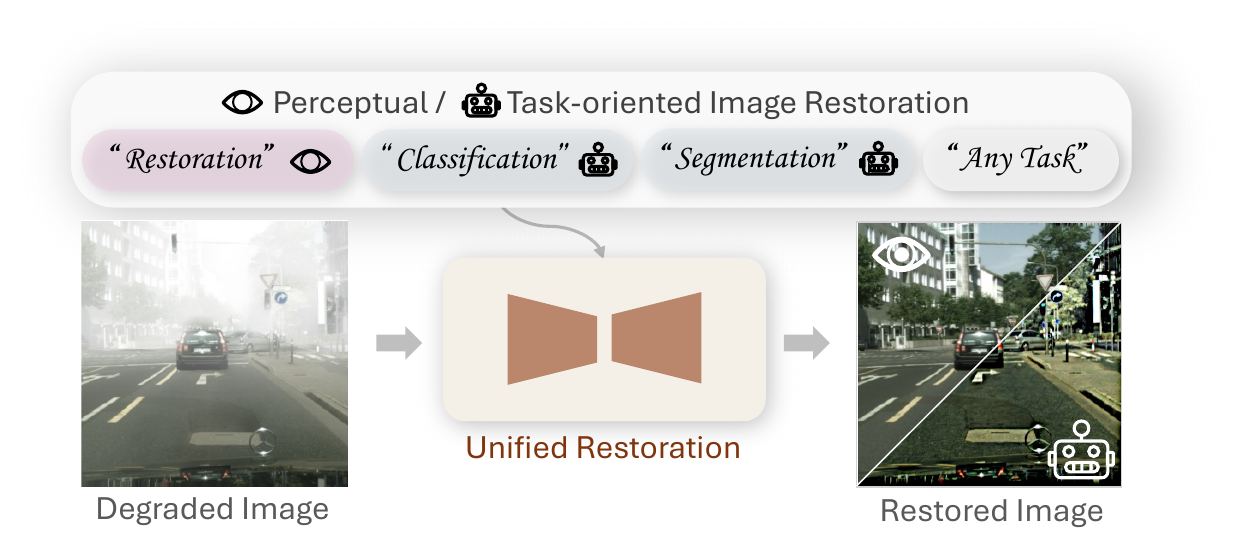}
        \caption{\method{}}
        \label{fig:teaser_uir}
    \end{subfigure}
    \caption{\textbf{Illustration of \method{}'s capabilities:} (a) PIR - Comparison with existing methods (\eg, URIE~\cite{son2020urie}, PromptIR~\cite{potlapalli2024promptir}) under adverse conditions. (b) TIR - Demonstrating \method{}'s robustness for downstream tasks such as classification and segmentation. (c) Unified Restoration - \method{}'s versatility in addressing PIR and TIR.}
    \label{fig:teasor}
\end{figure}

Perceptual image restoration (PIR) algorithms focus on improving visual clarity and fidelity of images by removing or reducing visible artifacts that affect their aesthetic quality. These methods focus on dealing with noise reduction~\cite{zamir2021multi,chihaoui2024self}, low-light enhancement~\cite{yi2023diff,zhang2023unified}, deblurring~\cite{nah2017deep,ren2023multiscale}, dehazing~\cite{chen2019pms, zamir2021multi}, deraining~\cite{wang2020model,yang2017deep}, and desnowing~\cite{wang2024ultra,chen2021all,liu2018desnownet}. While these algorithms can enhance the visual quality of images, they do not always improve performance in downstream tasks~\cite{huang2023counting,chen2022rvsl, chen2022sjdl}. This is because the factors contributing to visual quality often differ from those determining recognition quality~\cite{son2020urie,yang2023visual,pei2018does}.

On the other hand, task-oriented image restoration (TIR)~\cite{son2020urie,yang2023visual} is specifically designed to optimize images for applications that rely on computer vision, such as object detection~\cite{zhang2023adaptive,xu2021exploring}, classification~\cite{sharma2018classification,son2020urie}, and autonomous driving systems~\cite{saravanarajan2023improving,qian2024allweathernet}. This approach aligns the restoration process with the specific requirements of neural network models used in these applications, ensuring that the restored images are suitable for effective machine interpretation. Although these methods improve the performance of downstream tasks, they often produce results that are less visually appealing~\cite{zhang2023adaptive,liu2022image}.

Based on the analysis above, existing image restoration algorithms often face a trade-off, highlighting the challenge of balancing technical functionality with aesthetic quality. This dual functionality is crucial because real-world applications often require restoration processes to be specifically adapted to different scenarios. Thus, developing a model that can simultaneously enhance perceptual quality and improve performance for downstream tasks is essential. Such a unified image restoration framework can effectively perform across diverse settings, reducing system redundancy and boosting operational efficiency, as shown in~\figref{fig:teasor}.

In this paper, we propose a unified image restoration paradigm, \method{}, which simultaneously improves the performance of downstream tasks and the human perceptual quality of degraded images. \method{} leverages a diffusion prior~\cite{rombach2022high} as the backbone, recognized for its generative capabilities in producing high-quality images. However, these images are typically optimized for human aesthetics, which may not align with downstream task requirements. \method{} addresses this limitation by adapting the diffusion prior to meet both perceptual and functional needs, enabling the model to effectively bridge the gap between visual quality and downstream task performance.

To bridge this gap, we exploit the encoder features from the autoencoder within the diffusion model as complementary elements to tailor the diffusion prior to specific tasks. We introduce a Complementary Feature Restoration Module designed to reconstruct degraded features in the encoder. Subsequently, we propose a Task Feature Adapter, which harmonizes the diffusion features with the restored features within the decoder for various downstream tasks. Given the diversity of downstream tasks and the frequent necessity to adapt these tasks within existing models, the TFA module offers extendability to accommodate new tasks. Extensive experiments validate \method{}'s effectiveness, demonstrating enhancements in both PIR and TIR, with the potential for expansion to additional downstream tasks.

%
The contributions of this work are:
\begin{compactitem}
\item We introduce \method{}, a unified image restoration model that addresses both perceptual image restoration and task-oriented image restoration within a single framework. Experimental results show that \method{} surpasses existing methods in both visual quality and downstream task performance.

\item  We propose two components for \method{}: the CFRM and the TFA. These modules work together to adaptively complement the diffusion prior, enabling simultaneous restoration across diverse tasks.
\end{compactitem}

%% file: Related_Work.tex
\section{Related Work}
\subsection{Perceptual Image Restoration}
Perceptual image restoration aims to enhance the visual quality of images as perceived by humans, and it can be categorized into \textit{single} degradation and \textit{multiple} degradation restoration. Early work addressed \textit{single} degradation such as denoising~\cite{zamir2021multi,chihaoui2024self}, dehazing~\cite{dong2014learning, chen2019pms,zamir2021multi}, deraining~\cite{wang2020model,yang2017deep}, low-light enhancement~\cite{yi2023diff,zhang2023unified}, and deblurring~\cite{nah2017deep, ren2023multiscale}. To tackle \textit{multiple} degradations, methods like MPRNet~\cite{zamir2021multi} and NAFNet~\cite{chen2022simple} introduced unified solutions, while transformer-based approaches such as SwinIR~\cite{liang2021swinir} and Restormer~\cite{zamir2022restormer} gained traction for their versatility. Additionally, holistic approaches like All-in-One~\cite{li2020all}, TransWeather~\cite{valanarasu2022transweather} and PromptIR~\cite{potlapalli2024promptir} focused on improving visual quality across a wide range of conditions while providing enhanced adaptability and performance.
While these methods excel in enhancing visual quality, they do not always improve downstream vision tasks.

\subsection{Task-oriented Image Restoration}
Task-oriented image restoration aims to enhance downstream task performance, as studies~\cite{lee2022fifo, chen2022rvsl, yang2023visual, wang2020deep, yang2022self} show that image degradation significantly impairs high-level task accuracy. 
URIE~\cite{son2020urie} integrates image enhancement and recognition in an end-to-end framework to mitigate degradation effects. DIP~\cite{liu2022image} adapts image processing dynamically based on degradation factors. VDR-IR~\cite{yang2023visual} unifies semantic representations of diverse degraded images to recover intrinsic semantics effectively. While these methods enhance downstream task performance, they may result in images that are less visually pleasing.

\subsection{Diffusion Model}
Diffusion Models (DMs)~\cite{ho2020denoising} leverage a parameterized Markov chain to optimize the lower variational bound on the likelihood function, achieving state-of-the-art results in sample quality~\cite{rombach2022high, song2020denoising, song2020score, jiang2024scedit} and various applications~\cite{xu2023open, shao2022diffustereo, zheng2023diffuvolume, zhang2023adding}. 
Recently, DMs have become integral to image restoration, including super-resolution~\cite{saharia2022image, li2022srdiff}, inpainting~\cite{lugmayr2022repaint, rombach2022high, chung2022come}, and degradation restoration~\cite{ozdenizci2023restoring, zhu2023denoising, xia2023diffir}. StableSR~\cite{wang2024exploiting} effectively leverages diffusion priors for real-world super-resolution, resulting in superior reconstruction quality. DiffUIR~\cite{zheng2024selective} utilizes a specialized hourglass architecture to map degraded inputs to high-quality outputs, enhancing both global and local features effectively. DiffBIR~\cite{lin2023diffbir} follows a two-stage restoration strategy, first addressing specific degradations and then refining the image quality through a diffusion generation model. Despite these advances, adopting pre-trained diffusion models for both PIR and TIR remains an open challenge.


%% file: Methodology.tex
\section{Proposed Method}

\begin{figure*}
    \centering
    \includegraphics[width=0.98\linewidth]{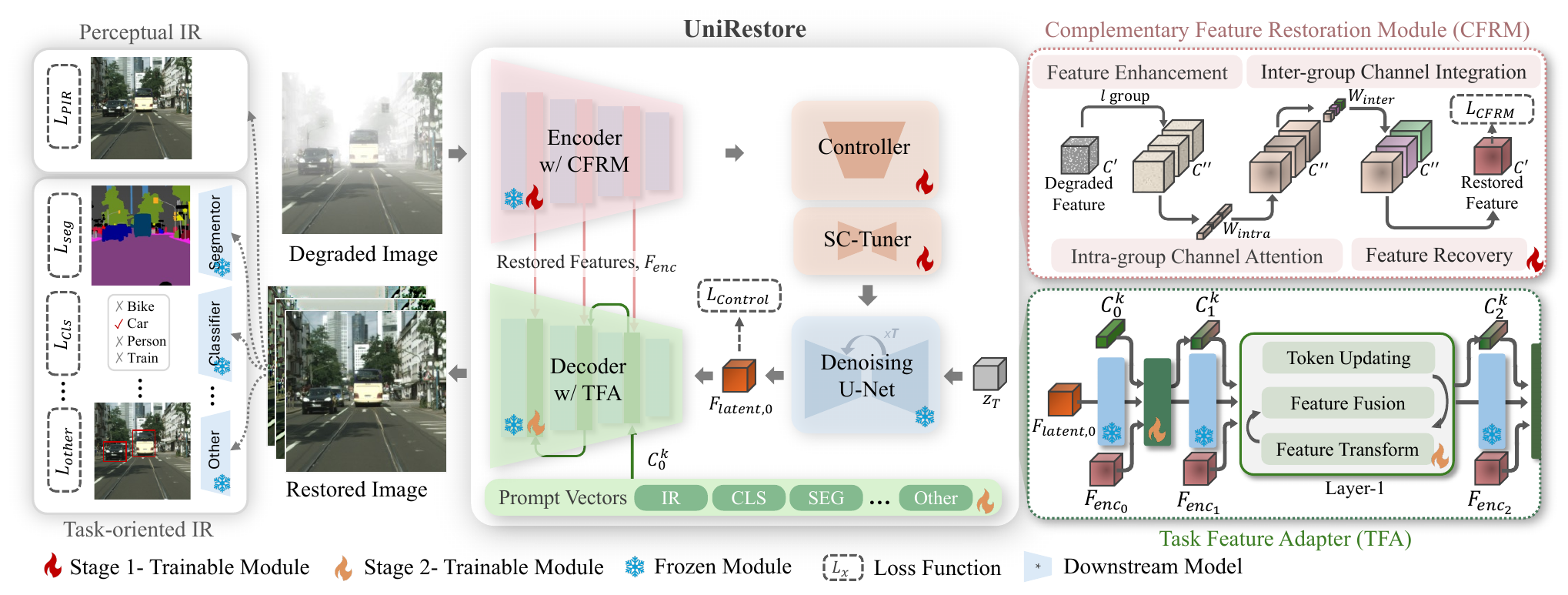}
    \caption{\textbf{Overview of \method{}.} \method{} augments the diffusion model by incorporating CFRMs and TFAs within the pre-trained autoencoder. The training process is divided into two stages: In the first stage, CFRM, Controller, and SC-Tuner are trained to restore clear encoder and latent features. In the second stage, the TFA is trained to adapt the restored encoder features and latent features for various downstream tasks, using task-specific prompts at the decoder to control the output restoration.}
    \label{fig:model_overview}
\end{figure*}

\subsection{Architecture of \method{}}
\method{} is built upon Stable Diffusion~\cite{rombach2022high}, leveraging its diffusion prior known for generating high-quality images. However, these images are optimized for human perception, which may not align with the requirements of machine vision tasks. To bridge this gap, we introduce two components: the Complementary Feature Restoration Module (CFRM) and the Task Feature Adapter (TFA). These modules adapt the diffusion prior to address diverse objectives, ensuring suitability for PIR and TIR tasks.

As illustrated in~\figref{fig:model_overview}, the input degraded image is processed through a modified encoder of VAE enhanced with the CFRM. The latent features produced by the final layer of this encoder are then fed into the Controller~\cite{zhang2023adding}, equipped with an SC-Tuner~\cite{jiang2024scedit}. The SC-Tuner, an enhanced module within the Controller architecture, integrates control signals efficiently with the Denoising U-Net. Subsequently, the noisy latent features are denoised to produce clear latent features, which are then passed to the decoder of the VAE augmented with the TFA. The restored features from the CFRM are input into the TFA, which adapts these features, enabling the decoder to generate outputs optimized for specific tasks.

\subsection{Complementary Feature Restoration Module}
The CFRM aims to restore and enhance features within the encoder, thereby providing complementary inputs to the decoder. As shown in ~\figref{fig:CFMRTFA}, the CFRM is integrated into the output of each encoder layer and consists of four steps:

\noindent\textbf{Feature Enhancement:}The enhancement begins with a NAFBlock~\cite{chu2022nafssr}, followed by a convolutional layer and group normalization~\cite{wu2018group}. Input features have dimensions $(C^{'}, H, W)$, where $C^{'}$ denotes the number of channels, $H$ the height, and $W$ the width. These features are expanded fourfold to dimensions $(4C^{'}, H, W)$ and then divided into $l$ groups, resulting in $(C^{''}, H, W)$, where $C^{''} = 4C^{'}/l$. 

\noindent\textbf{Intra-group Channel Attention:} In this stage, group convolution is employed to model learning from diverse degradation types. Initially, features are processed through a group convolution~\cite{krizhevsky2012imagenet} and Gate Linear Unit (GELU)~\cite{hendrycks2016gaussian}, followed by average pooling. Subsequently, a subsequent group convolution operation calculates the intra-group channel weights $W^i_\text{intra}$, each with dimensions $(C'', 1, 1)$, for groups indexed from 0 to $l$-1. This module sharpens the model's emphasis on pivotal intra-group features.

\noindent\textbf{Inter-group Channel Integration:} This module enhances contextual awareness by applying average pooling and a convolutional layer, yielding the inter-group channel weights $W_{inter}$ with dimension $(l, 1, 1)$.

\noindent\textbf{Feature Recovery:} In the final stage, a convolution layer merges the refined group features. These are then combined with the enhanced features via a skip connection.

The output feature of the CFRM serves as a complementary feature to the subsequent TFA.

\subsection{Task Feature Adapter}
The TFA leverages the restored features from the CFRM to adapt the original diffusion features for various objectives. The core idea is to integrate the CFRM output features at each layer with the corresponding output from the decoder at the same scale, enabling feature fusion for the targeted purpose. To achieve this, a straightforward approach involves designing a distinct feature adapter module for each task, which is then individually optimized using the relevant objectives and datasets. However, given the wide range of TIR tasks, this approach requires extensive model parameters and faces scalability challenges.

\begin{figure}[t!]
    \centering
    \begin{subfigure}{0.9\columnwidth}
        \centering
        \includegraphics[width=\linewidth]{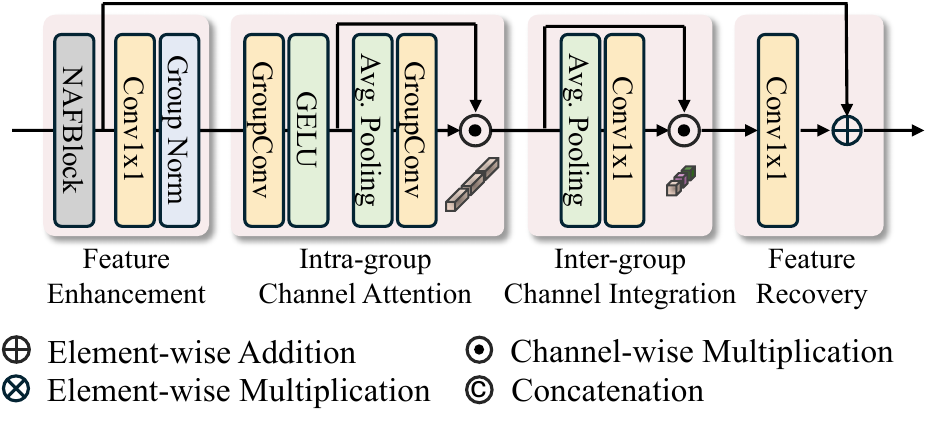}
        \caption{CFRM}
        \label{fig:module_CFRM}
    \end{subfigure}
    \hspace{-1.0cm} 
    \\
    \begin{subfigure}{0.9\columnwidth}
        \centering
        \includegraphics[width=\linewidth]{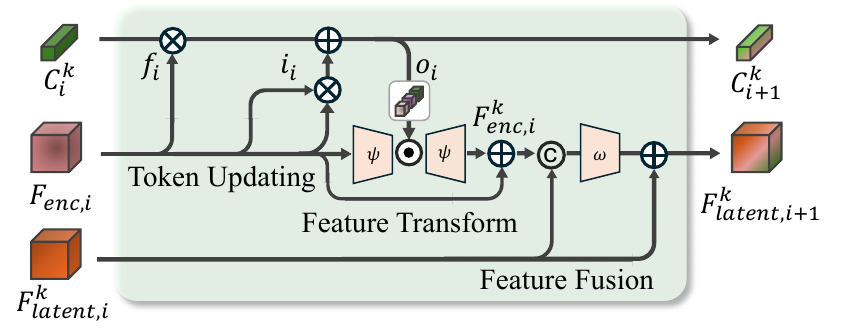}
        \caption{TFA}
        \label{fig:module_TFA}
    \end{subfigure}
    \caption{\textbf{Schematic diagrams of (a) the Complementary Feature Restoration Module and (b) the Task Feature Adapter.}}
    \label{fig:CFMRTFA}
\end{figure}

To address this limitation, we draw inspiration from prompt tuning~\cite{jia2022visual} and LSTM~\cite{hochreiter1997long} and propose an efficient architecture that reuses TFA, relying only on a lightweight, learnable prompt vector to adapt to different tasks effectively. As shown in \figref{fig:CFMRTFA}, for each task, we initialize a lightweight learnable prompt vector \(C_0^{k}\), where \(k\) represents the task index. This vector controls the weights of the CFRM features \(F_{\text{enc},i}\) in each decoder layer \(i\), dynamically combining them with the decoder's output features \(F_{\text{latent},i}^{k}\). This prompt is updated within layer \(i\) and passed to the next layer \(i+1\) as the input prompt for TFA. The procedure can be formulated as:
\begin{equation}
    \begin{split}
        f_i = \sigma \left( \vartheta \left(F_{\text{enc}, i}, \theta_{f, i} \right) \right) \\
        i_i = \sigma \left( \vartheta \left(F_{\text{enc}, i}, \theta_{i, i} \right) \right) \\
        C_{i+1}^k = f_i \otimes C_i^k + i_i \otimes \tanh \left( \vartheta \left(F_{\text{enc}, i}, \theta_{c, i} \right) \right) \\
        o_i = \tanh ( \xi (C_{i+1}^k, \theta_{o, i})) \\
        F^{k}_{\text{enc}, i} = \psi(F_{\text{enc}, i}, o_i, \theta_{t, i}) + F_{\text{enc}, i} \\
        F^{k}_{\text{latent}, i+1} = \omega ((F^{k}_{\text{enc}, i}, F^{k}_{\text{latent}, i}), \theta_{l, i}) + F^{k}_{\text{latent}, i}
    \end{split}
\label{eq:lstm}
\end{equation}
where $\sigma$ and $\tanh$ represent the softmax and $\tanh$ activate function, $\otimes$ denotes the element-wise multiplication. $\vartheta \left(\cdot, \theta_{x, i} \right)$ is the prompt updating project function, involving instance normalization, convolution operation, GELU, and global average pooling. $\xi(\cdot, \theta_{o, i}))$ and $\omega(\cdot, \theta_{l, i}))$ are simple projection layers. $\psi(\cdot,c,\theta_{o, i})$ is a tuner-operator \cite{jiang2024scedit} with a channel attention based on condition $c$. Through this process, the restored features adapt to the diffusion features at each scale through different prompts, ultimately producing images suitable for downstream tasks.

\subsection{Training Pipeline}
\label{sec:twostage}
Our training pipeline consists of three stages:

\noindent\textbf{Stage 1} primarily focuses on adapting stable diffusion for the image restoration context. In this stage, we utilize the PIR dataset to train the CFRM, Controller~\cite{zhang2023adding}, and SC-Tuner~\cite{jiang2024scedit}. The loss function, \(\mathcal{L}_{\text{CFRM}}\), is designed to enable the CFRM to learn the restoration of degraded features to their clear states. Specifically, clear latent features \(f_i^{\text{Clear}}\) are extracted from the $i^{\text{th}}$ layer of the vanilla encoder using a clean image. Similarly, restored latent features \(f_i^{\text{Restored}}\) are derived by inputting a degraded image into the encoder integrated with the CFRM. The loss function is defined as:
\begin{equation}
\mathcal{L}_{\text{CFRM}} = \sum_{i=1}^M \lambda_i (f_i^{\text{Clear}} - f_i^{\text{Restored}}),
\end{equation}
where \(M\) denotes the number of layers in the encoder, and \(\lambda_i\) represents the scaling weight for the \(i^{\text{th}}\) layer.

Additionally, the Controller and SC-Tuner are trained using the loss function \(\mathcal{L}_{\text{Control}}\). This function is designed to align the clear latent features \(z_0\) of the clear image with the reconstructed latent features \(\hat{z}^{t}_0\) at any given sampled step \(t\). The loss function can be expressed as:
\begin{equation}
\mathcal{L}_{\text{Control}} = \|z_0 - \hat{z}^{t}_0\|_2^2.
\end{equation}
The total loss at Stage 1 is:
\begin{equation}
\mathcal{L}_{\text{Stage 1}} = \mathcal{L}_{\text{CFRM}} + \mathcal{L}_{\text{Control}}.
\end{equation}
During this stage, TFA is not integrated into the decoder. 

\vspace{1mm}
\noindent\textbf{Stage 2} aims to optimize TFA to adapt the diffusion prior to different objectives. Therefore, CFRM, Controller, and SC-Tuner do not undergo parameter updates during this stage. 

We optimize the network using objectives specific to each task, defined as:
\begin{equation}
\mathcal{L}_{\text{Stage 2}} = \sum_{i=1}^N \beta_{\text{Task}}^{i} \mathcal{L}_{\text{Task}}^{i},
\end{equation}
where \(N\) represents the number of tasks, and \(\beta_{\text{Task}}^{i}\) are the weighting coefficients that adjust the importance of each task-specific loss \(\mathcal{L}_{\text{Task}}^{i}\) on the overall multi-task learning objective. 

In this paper, we aim to address both PIR and TIR tasks simultaneously, selecting semantic segmentation and image classification as representative TIR tasks. For semantic segmentation, we employ cross-entropy loss using segmentation labels, while for classification, we also use cross-entropy loss but with class labels. For the PIR task, Mean Squared Error is applied to compare the reconstructed images against their corresponding ground truths.

The overall loss for Stage 2 of \method is formulated as:
\begin{equation}
\mathcal{L}_{\text{Stage 2}} = \beta_{\text{PIR}} \mathcal{L}_{\text{PIR}} + \beta_{\text{Seg}} \mathcal{L}_{\text{Seg}} + \beta_{\text{Cls}} \mathcal{L}_{\text{Cls}}
\label{eq:total_loss}
\end{equation}
These tasks may originate from different datasets, and we distribute images from these tasks across each batch. Losses are calculated based on the input and its corresponding task before the model parameters are updated.

\noindent\textbf{Introducing Additional Tasks.} After training \method{} with the two-stage process, adding more tasks in TIR requires only the introduction of a new task-specific prompt, which can then be optimized with the corresponding objective and training data. This process does not require data or loss functions from the original tasks, as only the new task-specific prompt is updated.


%% file: Implementation_Detail.tex
\section{Implementation Details}
\label{sec:implementation}
To evaluate the effectiveness of the proposed \method{}, experiments are conducted in PIR and TIR. Within TIR, image classification and semantic segmentation are chosen as downstream tasks. Detailed descriptions of the implementation details are provided in the Supplementary Material.

\subsection{Training Dataset}
We reference the dataset configurations from previous PIR~\cite{xia2023diffir} and TIR~\cite{son2020urie} studies. Specifically, we use a blend of the DIV2K~\cite{agustsson2017ntire}, Flickr2K~\cite{Lim_2017_CVPR_Workshops}, and OST~\cite{wang2018recovering} datasets for PIR tasks. For image classification, we randomly select 80,000 images from the training set of ImageNet~\cite{deng2009imagenet}, and for semantic segmentation, we use the training set from the Cityscapes datasets~\cite{cordts2016cityscapes}. These datasets are synthesized with 15 types of degradation, including blur, noise, adverse weather conditions, etc., following the procedures outlined in~\cite{hendrycks2019benchmarking} to create our training set.

\subsection{Evaluation Dataset}
For PIR evaluation, \method{} is evaluated on the test set of DIV2K~\cite{agustsson2017ntire} using the same degradation synthesis procedure as in training. To further assess the robustness of \method{} on unseen data, we utilize multiple benchmarks with synthetic degradations. These benchmarks include various tasks such as image denoising, which uses a composite dataset labeled 'Noise'—comprising Urban100~\cite{huang2015single}, BSD68~\cite{martin2001database}, CBSD68~\cite{martin2001database}, Kodak~\cite{martin2001database}, McMaster~\cite{martin2001database}, and Set12~\cite{martin2001database}. Additionally, we utilize Rain100L~\cite{yang2017deep} for deraining, RESIDE~\cite{li2018benchmarking} for dehazing, UHDSnow~\cite{wang2024ultra} for desnowing, and GoPro~\cite{nah2017deep} for deblurring.

In the TIR context, \method{} is evaluated as follows: For classification, we sample 20,000 images from the test set ImageNet and utilize the entire CUB dataset~\cite{wah2011caltech} for validation of unseen data, applying the same degradation synthesis method as during training. For semantic segmentation, \method{} is assessed on the test set of Cityscapes using identical degradation synthesis. Additionally, \method{} is tested on the FoggyCityscapes dataset~\cite{sakaridis2018semantic}, specifically the subsets Fog1, Fog2, and Fog3, with results reported as the average across these three subsets. Further evaluations include the unseen ACDC dataset~\cite{sakaridis2021acdc}.




\subsection{Evaluation Protocol}
For the evaluation of PIR, we utilize Peak Signal-to-Noise Ratio (PSNR) and Structural Similarity Index Measure (SSIM). For image classification, we measure performance using accuracy (ACC), and for semantic segmentation, we use the mean Intersection over Union (mIoU).

%% file: Experimental_Results.tex
\begin{table*}[ht!]
    \centering
    \scalebox{0.75}{
        \begin{tabular}{@{}l|cc|cc|cc|cc|cc|cc|cc@{}}
            \toprule[1.3pt]            
& \multicolumn{2}{c|}{\textit{Seen Dataset}}  & \multicolumn{10}{c|}{\textit{Unseen Datasets}} &  \multicolumn{2}{c}{\multirow{2}{*}{\textit{Average}}} \\
\multirow{2}{*}{\centering Methods}& \multicolumn{2}{c|}{DIV2K \cite{agustsson2017ntire}}  & \multicolumn{2}{c|}{Rain100L \cite{yang2017deep}} & \multicolumn{2}{c|}{RESIDE \cite{li2018benchmarking}} & \multicolumn{2}{c|}{UHDSnow \cite{wang2024ultra}} & \multicolumn{2}{c|}{Noise \cite{martin2001database, huang2015single}} & \multicolumn{2}{c|}{GoPro \cite{nah2017deep}} & \multicolumn{2}{c}{} \\ 
\cmidrule[0.8pt]{2-15}
 & PSNR$\uparrow$  & SSIM$\uparrow$ & PSNR$\uparrow$  & SSIM$\uparrow$ & PSNR$\uparrow$  & SSIM$\uparrow$ & PSNR$\uparrow$  & SSIM$\uparrow$ & PSNR$\uparrow$  & SSIM$\uparrow$ & PSNR$\uparrow$  & SSIM$\uparrow$ & PSNR$\uparrow$ & SSIM$\uparrow$ \\

            \midrule
			DIP \cite{liu2022image}                  & 18.47 & 0.5810 & 22.65 & 0.7884 & 21.30 & 0.7819 & 19.03 & 0.8089 & 15.41 & 0.2494 & 23.08 & 0.8041 & 17.13 & 0.5734 \\
			DIP* \cite{liu2022image}                 & 18.62 & 0.5516 & 23.16 & 0.8097 & 19.83 & 0.7586 & 16.77 & 0.7830 & 14.51 & 0.2328 & 21.05 & 0.7624 & 16.28 & 0.5569 \\
			URIE \cite{son2020urie}                  & 17.72 & 0.5202 & 20.97 & 0.7293 & 18.30 & 0.7449 & 18.11 & 0.7626 & 18.57 & 0.5180 & 19.21 & 0.5683 & 18.81 & 0.6406 \\
			URIE* \cite{son2020urie}                 & 17.98 & 0.5967 & 19.97 & 0.6993 & 20.37 & 0.7694 & 16.18 & 0.7526 & 17.41 & 0.3624 & 18.57 & 0.4624 & 18.41 & 0.6071 \\
            NAFNet \cite{chen2022simple}             & 22.23 & 0.7905 & 24.57 & 0.8178 & 25.13 & 0.8632 & 20.71 & 0.8672 & 23.22 & 0.6951 & 22.18 & 0.8042 & 23.01 & 0.8063 \\
			NAFNet* \cite{chen2022simple}            & 19.81 & 0.7005 & 20.51 & 0.7314 & 21.24 & 0.8178 & 18.39 & 0.7958 & 20.38 & 0.6019 & 19.79 & 0.7293 & 20.02 & 0.7295 \\
			PromptIR \cite{potlapalli2024promptir}   & \underline{23.90} & 0.8321 & 28.17 & 0.9034 & \underline{27.26} & \underline{0.8957} & \underline{22.10} & \underline{0.8877} & 23.72 & 0.7269 & \underline{23.93} & 0.8221 & \underline{24.85} & 0.8447 \\
			PromptIR* \cite{potlapalli2024promptir}  & 21.94 & 0.7421 & 24.76 & 0.8134 & 24.16 & 0.8317 & 19.13 & 0.8265 & 19.68 & 0.6283 & 20.18 & 0.7657 & 21.64 & 0.7680 \\
			DiffBIR \cite{lin2023diffbir}            & 22.76 & 0.8053 & 27.25 & 0.8695 & 26.97 & 0.8770 & 20.84 & 0.8785 & 23.67 & \underline{0.7661} & 23.49 & 0.8076 & 24.16 & 0.8340 \\
			DiffBIR* \cite{lin2023diffbir}           & 18.32 & 0.6847 & 23.48 & 0.8143 & 23.13 & 0.8068 & 18.29 & 0.8167 & 21.59 & 0.6419 & 20.13 & 0.7413 & 20.82 & 0.7510 \\
			DiffUIR \cite{zheng2024selective}        & 23.79 & \underline{0.8397} & \underline{28.25} & \underline{0.9154} & 27.12 & 0.8820 & 20.74 & 0.8753 & \underline{24.27} & 0.7481 & \underline{23.93} & \underline{0.8241} & 24.68 & \underline{0.8474} \\
			DiffUIR* \cite{zheng2024selective}       & 21.47 & 0.7742 & 25.44 & 0.8276 & 23.58 & 0.8174 & 18.62 & 0.8318 & 22.76 & 0.6691  & 21.71 & 0.7649 & 22.26 & 0.7808 \\
			\rowcolor{LightCyan} \textbf{\method{}}  & \textbf{24.32} & \textbf{0.8434} & \textbf{30.02} & \textbf{0.9237} & \textbf{27.91} & \textbf{0.9043} & \textbf{23.44} & \textbf{0.8943} & \textbf{24.37} & \textbf{0.7811} & \textbf{25.94} & \textbf{0.8541} & \textbf{26.00} & \textbf{0.8668} \\
            \bottomrule[1.3pt]
        \end{tabular}}
    \caption{\textbf{Performance comparison of existing methods on one seen and five unseen PIR datasets.}}
    \label{tab:pir_performance}
\end{table*}

\begin{figure*}[t!]
  \centering
  \setlength{\tabcolsep}{1pt}
    \footnotesize
    \begin{tabular}{cccccc}
        \includegraphics[width=.16\textwidth, height=2.3cm]{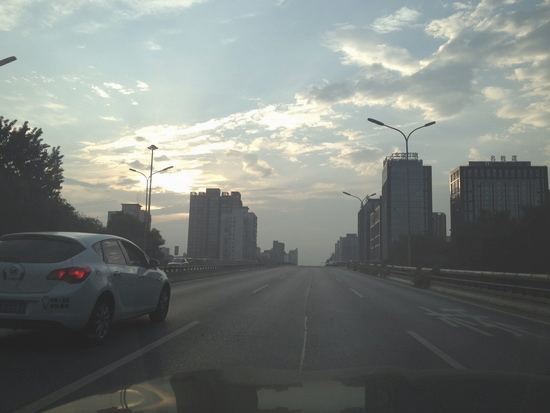} &
        \includegraphics[width=.16\textwidth, height=2.3cm]{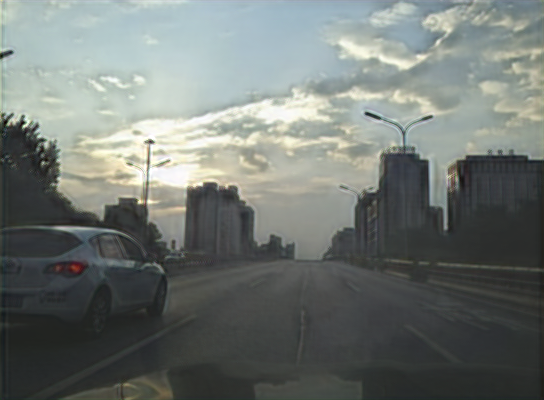} &
        \includegraphics[width=.16\textwidth, height=2.3cm]{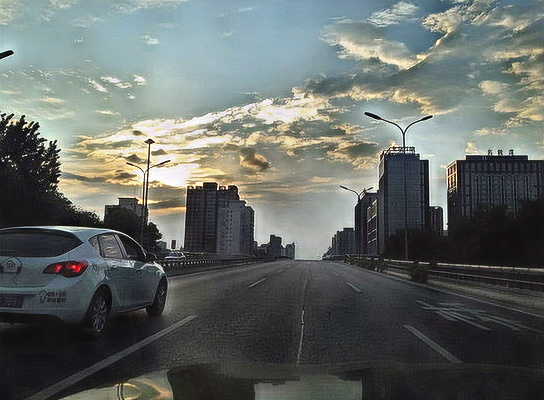} &
        \includegraphics[width=.16\textwidth, height=2.3cm]{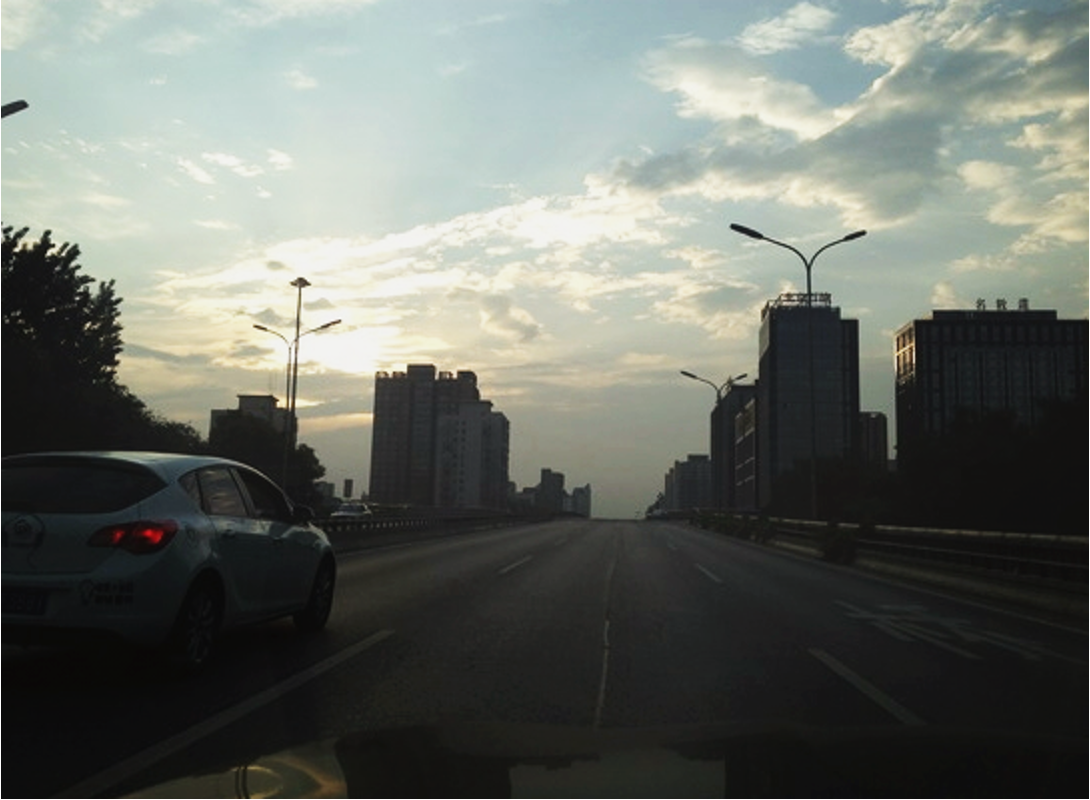} &
        \includegraphics[width=.16\textwidth, height=2.3cm]{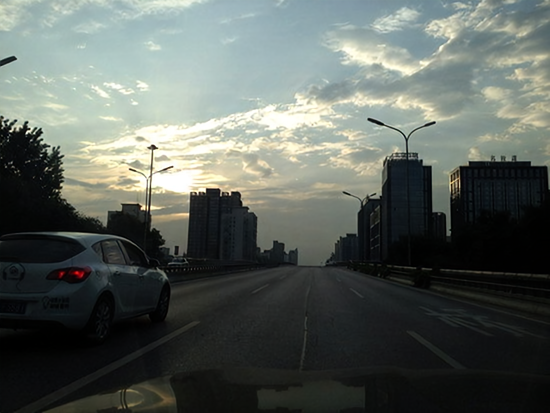} &
        \includegraphics[width=.16\textwidth, height=2.3cm]{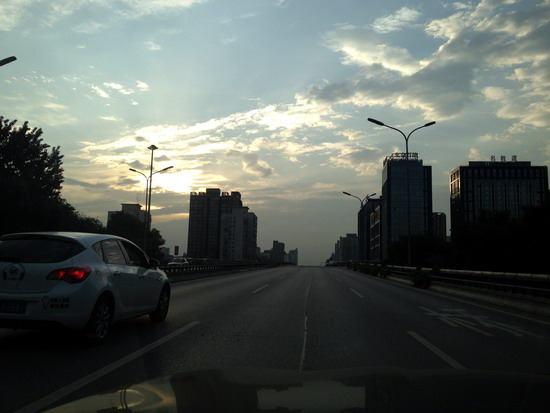} \\
        \includegraphics[width=.16\textwidth, height=2.3cm]{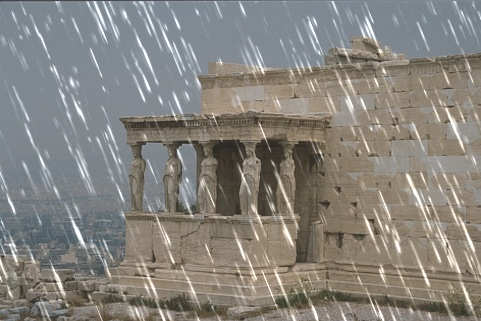} &
        \includegraphics[width=.16\textwidth, height=2.3cm]{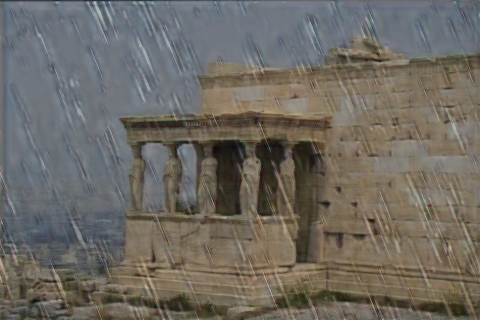} &
        \includegraphics[width=.16\textwidth, height=2.3cm]{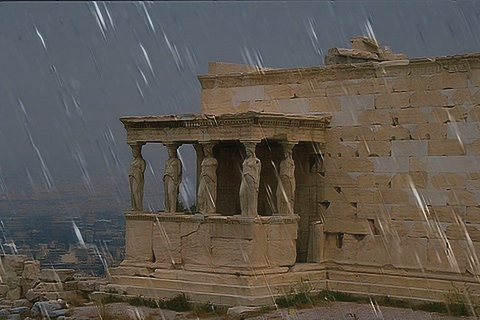} &
        \includegraphics[width=.16\textwidth, height=2.3cm]{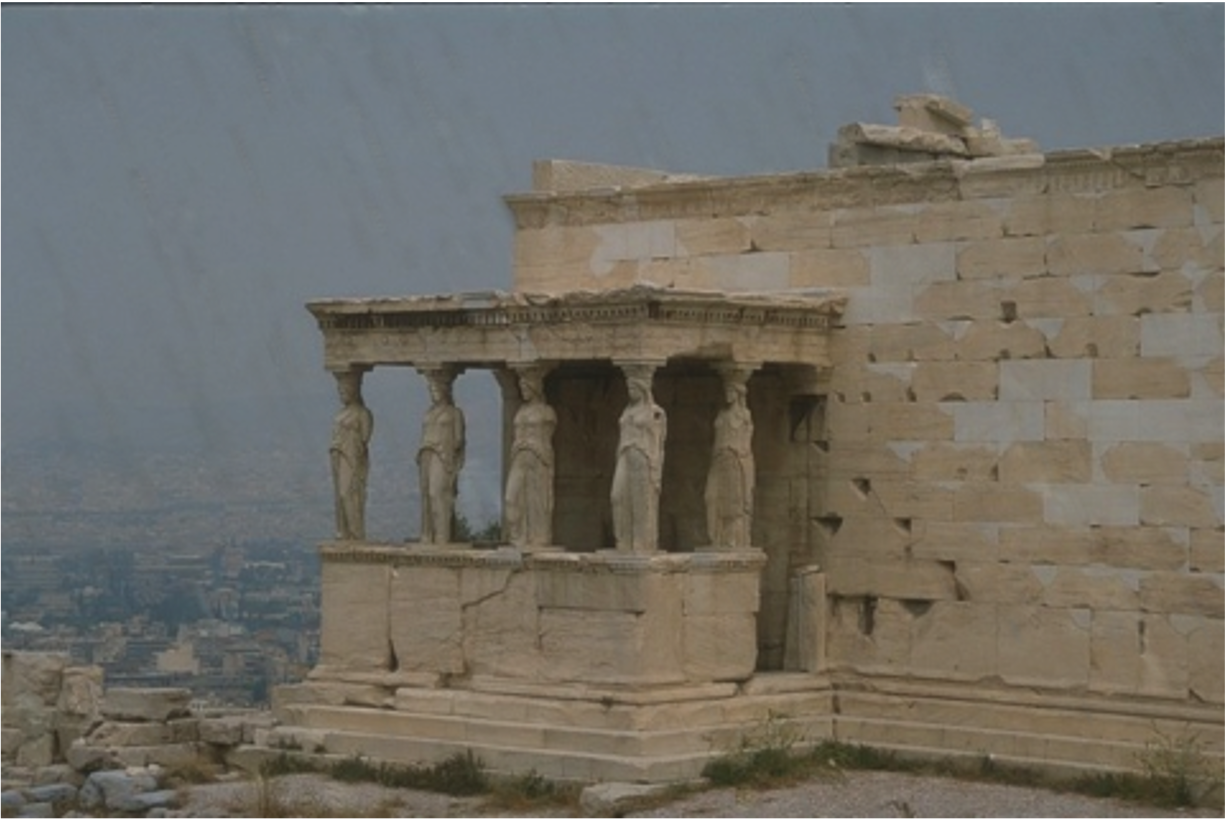} &
        \includegraphics[width=.16\textwidth, height=2.3cm]{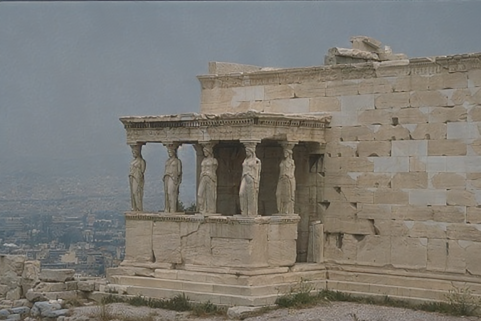} &
        \includegraphics[width=.16\textwidth, height=2.3cm]{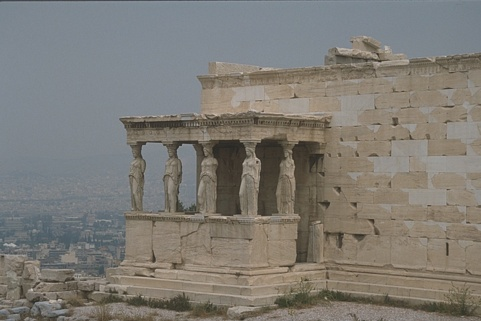} \\
        \includegraphics[width=.16\textwidth, height=2.3cm]{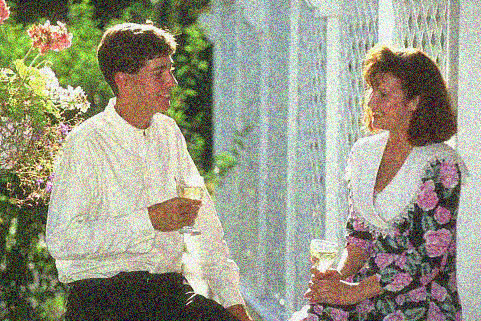} &
        \includegraphics[width=.16\textwidth, height=2.3cm]{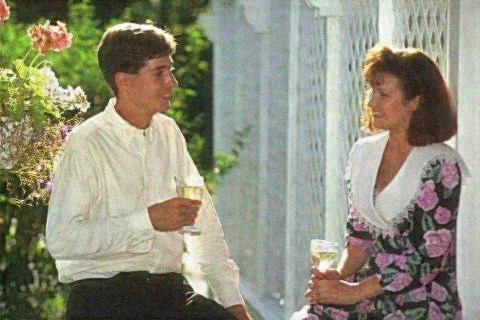} &
        \includegraphics[width=.16\textwidth, height=2.3cm]{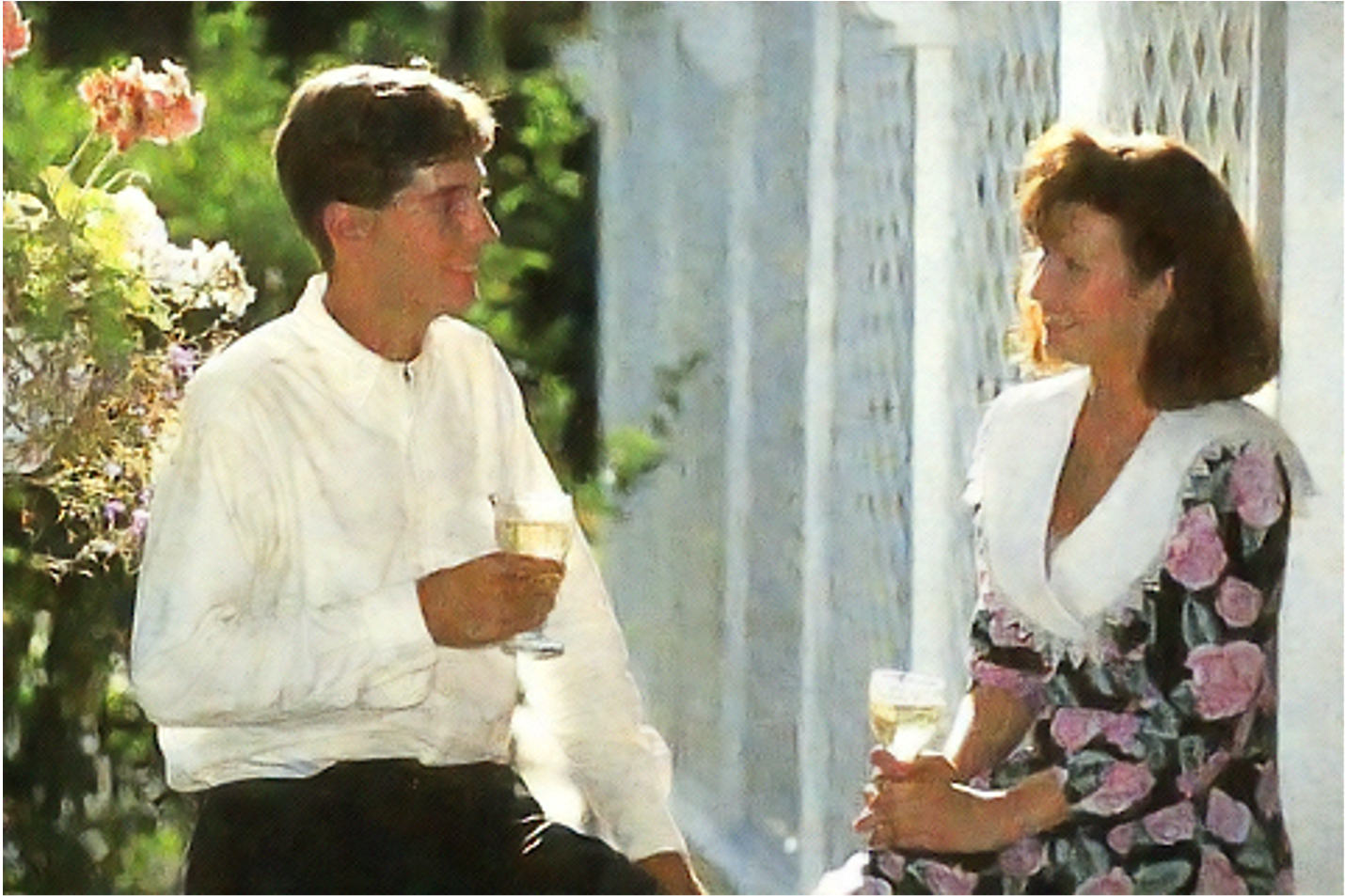} &
        \includegraphics[width=.16\textwidth, height=2.3cm]{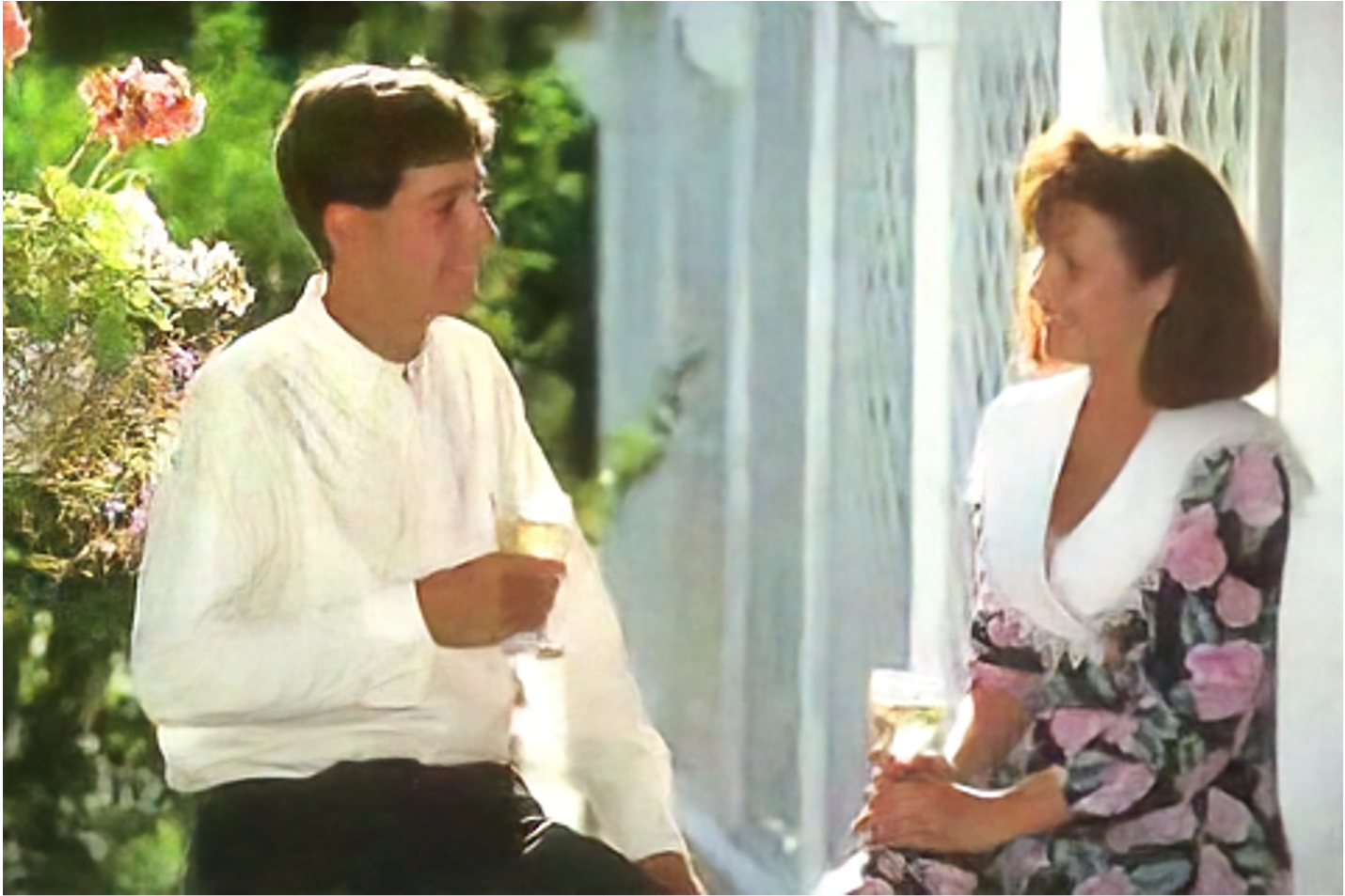} &
        \includegraphics[width=.16\textwidth, height=2.3cm]{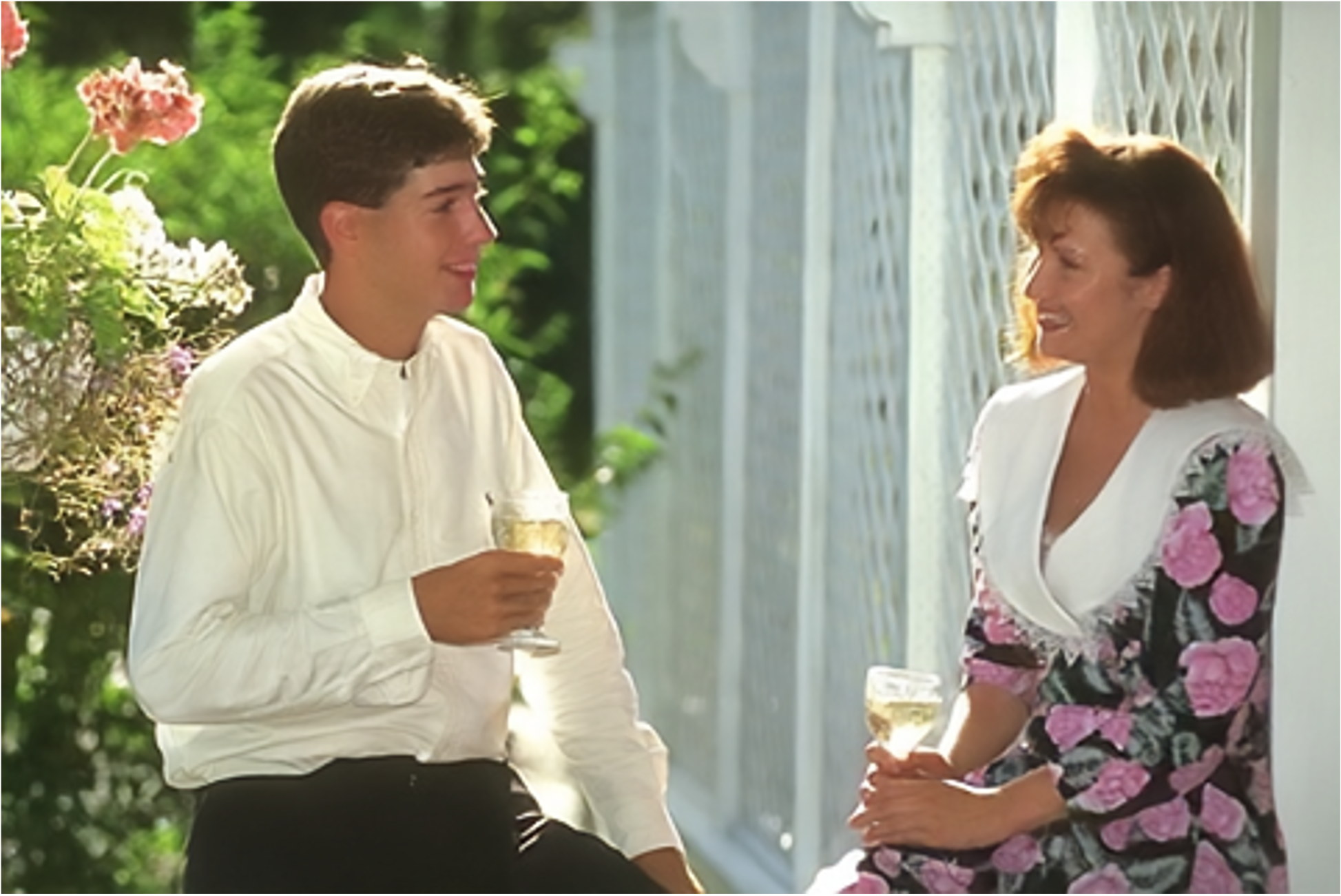} &
        \includegraphics[width=.16\textwidth, height=2.3cm]{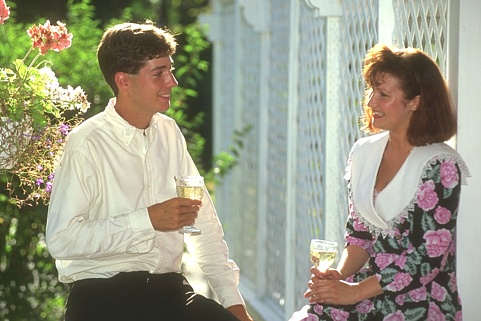} \\
        LQ & URIE & PromptIR & DiffUIR & \method{} & HQ  \\
    \end{tabular}
    \caption{\textbf{Qualitative analysis of perceptual image restoration:} A visual comparison on unseen datasets highlighting the performance improvements of the \method{} over existing methods.}
    \label{fig:pir_vis}
\end{figure*}

\section{Experiments}\label{sec:exp}
\subsection{Baselines}
To evaluate the effectiveness of \method{}, we compare it against multiple TIR methods, including DIP~\cite{liu2022image} and URIE~\cite{son2020urie}, as well as PIR methods such as NAFNet~\cite{chen2022simple}, and PromptIR~\cite{potlapalli2024promptir}. Furthermore, we include comparisons with diffusion-based approaches including DiffBIR~\cite{lin2023diffbir} and DiffUIR~\cite{zheng2024selective}. We report results across two settings: First, models are trained only for their intended purpose (\textit{i.e.}, PIR or TIR) using the corresponding dataset from our training set. Second, for a fair comparison and following the training pipeline of \method{}, all baseline models are initially trained on the PIR training set and then fine-tuned on multiple downstream tasks (\textit{i.e.}, PIR and TIR) using the loss function in \eqref{eq:total_loss}, indicated by the suffix ''*''.



\subsection{Perceptual Image Restoration}
The results presented in \tabref{tab:pir_performance} show that \method{} achieves the best overall performance on the seen dataset. Additionally, it highlights \method{}'s generalizability across several unseen datasets, especially in high-resolution scenarios (UHDSnow) and dynamic scenes (GoPro). Moreover, TIR methods generally show limited performance in PIR task, as their learning objectives are optimized for specific downstream tasks. Additionally, in scenarios involving multiple downstream tasks, both PIR and TIR methods exhibit limited performance in due to the absence of a mechanism to learn different objectives simultaneously. We also present a qualitative comparison in~\figref{fig:pir_vis}, where \method{} reconstructs more details and delivers better visual quality.

\begin{table}[t!]
    \centering
    \setlength\tabcolsep{3pt}
    \scalebox{0.65}{
        \begin{tabular}{l|cc|cc} 
            \toprule[1.3pt]
			\multirow{3}{*}{Inputs} & \multicolumn{2}{c|}{Seen Dataset}          & \multicolumn{2}{c}{Unseen Dataset} \\ 

			              & \multicolumn{2}{c|}{ImageNet \cite{deng2009imagenet}} & \multicolumn{2}{c}{CUB \cite{wah2011caltech}} \\ 
            \cmidrule[0.8pt]{2-5}
             & ResNet-50~\cite{he2016deep} $\uparrow$         & ViT-B~\cite{dosovitskiy2020image} $\uparrow$                 &  ResNet-50~\cite{he2016deep} $\uparrow$ & ViT-B~\cite{dosovitskiy2020image} $\uparrow$ \\ 
            \midrule
            \textit{LQ}                & \textit{51.75} & \textit{67.65} & \textit{33.69} & \textit{44.83 } \\
            \hdashline
            DIP \cite{liu2022image}                         & 61.55 & 72.05 & 47.91 & 54.10 \\
            DIP* \cite{liu2022image}                        & 59.80 & 70.35 & 45.99 & 52.48 \\
            URIE \cite{son2020urie}                         & \underline{66.65} & \underline{73.95} & \underline{49.64} & 57.24 \\
            URIE* \cite{son2020urie}                        & 65.20 & 72.15 & 46.89 & 54.93 \\
            NAFNet \cite{chen2022simple}                    & 60.35 & 70.80 & 46.47 & 53.82 \\
            NAFNet* \cite{chen2022simple}                   & 57.65 & 68.25 & 43.17 & 51.88 \\
            PromptIR \cite{potlapalli2024promptir}          & 65.25 & 73.90 & 49.52 & \underline{58.04} \\
            PromptIR* \cite{potlapalli2024promptir}         & 64.05 & 73.00 & 48.52 & 57.39 \\
            DiffBIR \cite{lin2023diffbir}                   & 59.30 & 68.05 & 41.68 & 52.38 \\
            DiffBIR* \cite{lin2023diffbir}                  & 57.55 & 66.85 & 40.65 & 51.34 \\
            DiffUIR \cite{zheng2024selective}               & 62.35 & 72.10 & 46.75 & 57.28 \\
            DiffUIR* \cite{zheng2024selective}              & 61.15 & 71.60 & 45.44 & 56.31 \\
            \rowcolor{LightCyan} \textbf{\method{}} & \textbf{71.65} & \textbf{77.05} & \textbf{53.70} & \textbf{60.79} \\
            \hdashline
            \textit{HQ} & \textit{72.80} & \textit{78.70} & \textit{58.22} & \textit{64.41} \\
            \bottomrule[1.3pt]
        \end{tabular}
    }
\caption{\textbf{Performance comparison of existing methods on seen and unseen datasets for image classification.} Results are reported in terms of accuracy.}
    \label{tab:cls_performance}
\end{table}

\begin{figure}[t!]
  \centering
  \setlength{\tabcolsep}{1pt}
  \footnotesize
    \begin{tabular}{cccccc}
        \includegraphics[width=.09\textwidth, height=1.4cm]{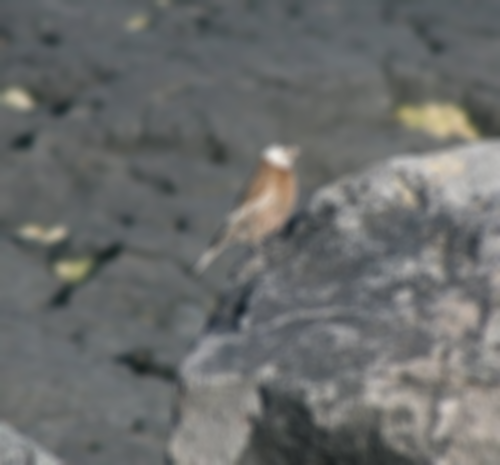} &
        \includegraphics[width=.09\textwidth, height=1.4cm]{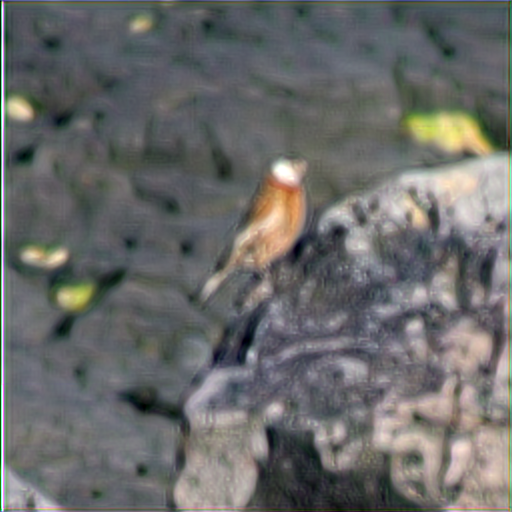} &
        \includegraphics[width=.09\textwidth, height=1.4cm]{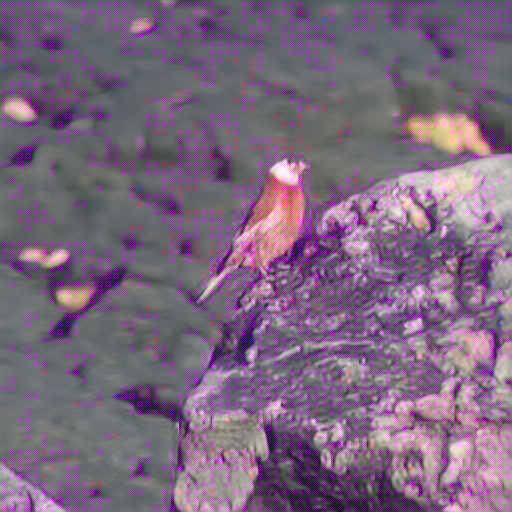} &
        \includegraphics[width=.09\textwidth, height=1.4cm]{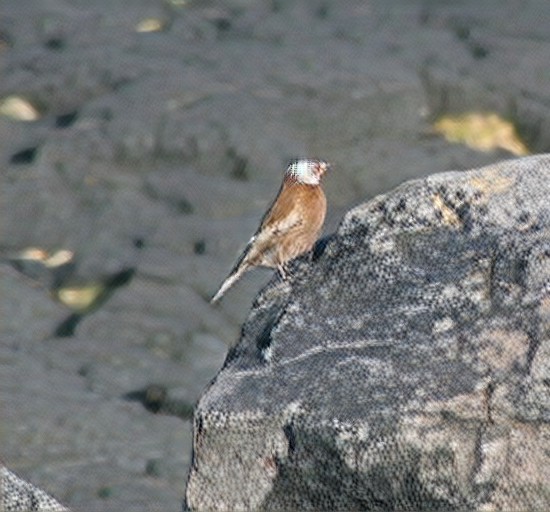} &
        \includegraphics[width=.09\textwidth, height=1.4cm]{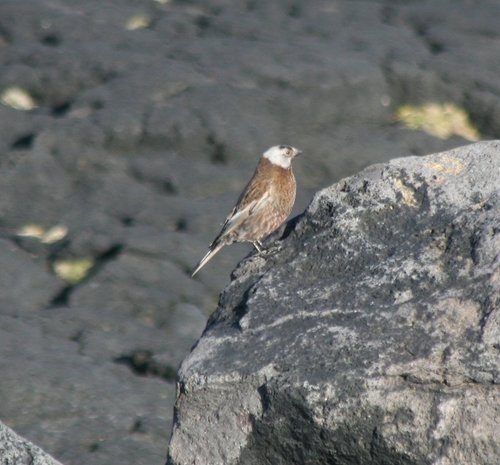} \\

        \includegraphics[width=.09\textwidth, height=1.4cm]{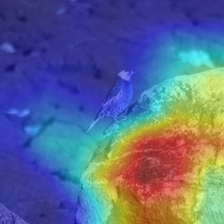} &
        \includegraphics[width=.09\textwidth, height=1.4cm]{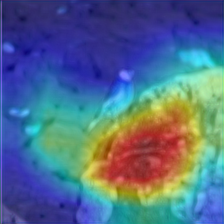} &
        \includegraphics[width=.09\textwidth, height=1.4cm]{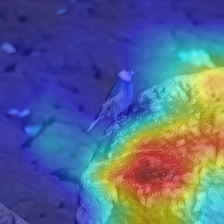} &
        \includegraphics[width=.09\textwidth, height=1.4cm]{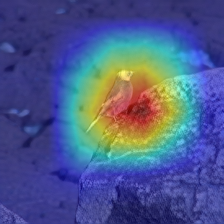} &
        \includegraphics[width=.09\textwidth, height=1.4cm]{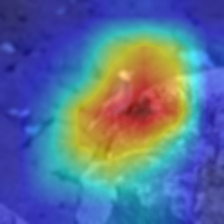} \\

        \includegraphics[width=.09\textwidth, height=1.4cm]{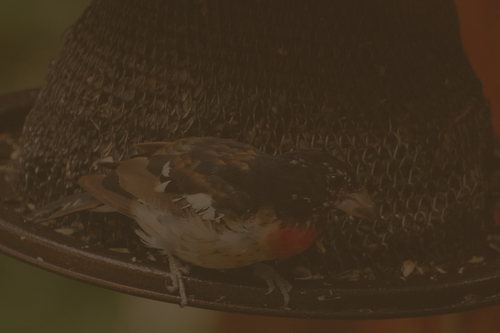} &
        \includegraphics[width=.09\textwidth, height=1.4cm]{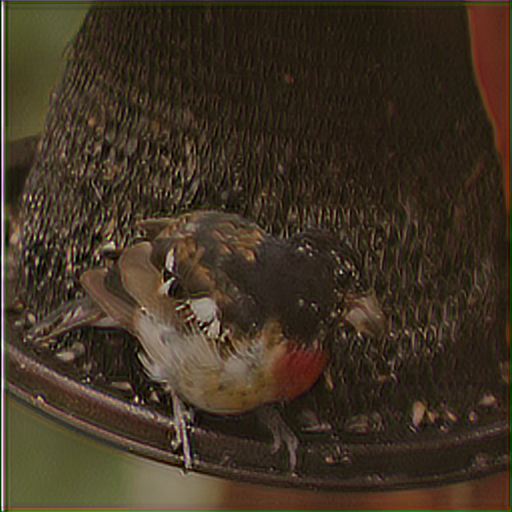} &
        \includegraphics[width=.09\textwidth, height=1.4cm]{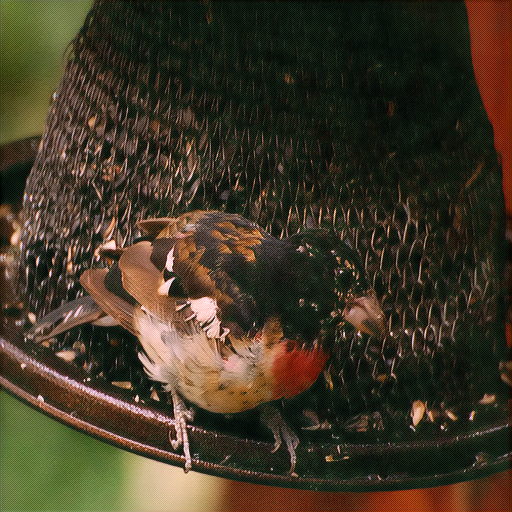} &
        \includegraphics[width=.09\textwidth, height=1.4cm]{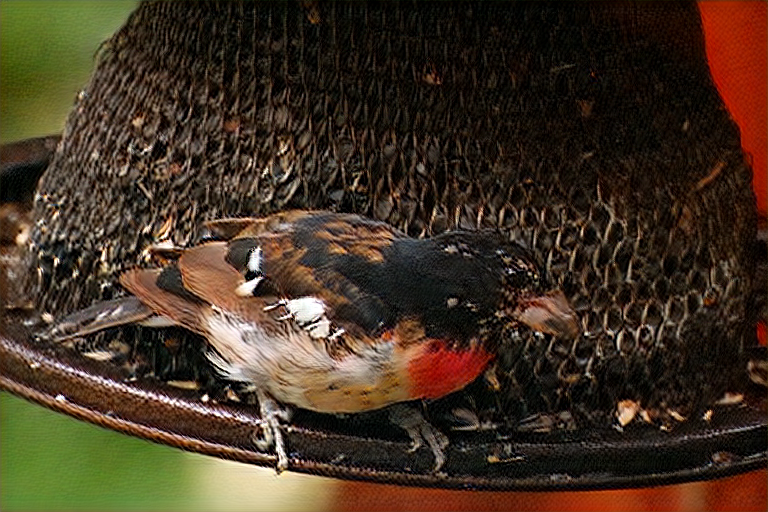} &
        \includegraphics[width=.09\textwidth, height=1.4cm]{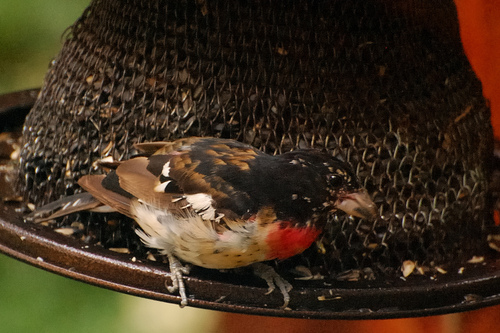} \\

        \includegraphics[width=.09\textwidth, height=1.4cm]{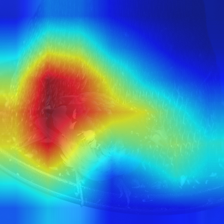} &
        \includegraphics[width=.09\textwidth, height=1.4cm]{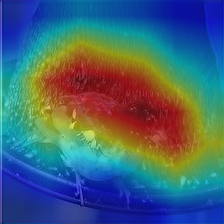} &
        \includegraphics[width=.09\textwidth, height=1.4cm]{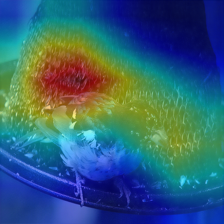} &
        \includegraphics[width=.09\textwidth, height=1.4cm]{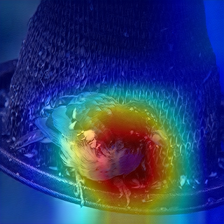} &
        \includegraphics[width=.09\textwidth, height=1.4cm]{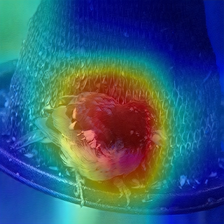} \\
        LQ & URIE & PromptIR & \method{} & HQ  \\
    \end{tabular}
    \caption{\textbf{Qualitative analysis of image classification.} The first and third rows display the input images, while the second and fourth rows show their corresponding activation maps.}
    \label{fig:cls_vis}
\end{figure}

\begin{table}[t!]
    \centering
    \setlength\tabcolsep{1pt}
    \scalebox{0.65}{
        \begin{tabular}{l|cc|c|c} 
            \toprule[1.3pt]
            \multirow{3}{*}{Inputs} & \multicolumn{3}{c|}{Seen Dataset} & \multicolumn{1}{c}{Unseen Dataset} \\
                    & \multicolumn{2}{c|}{Cityscapes \cite{sakaridis2018semantic}}                & \multicolumn{1}{c|}{FoggyCityscapes \cite{sakaridis2018semantic}} & \multicolumn{1}{c}{ACDC \cite{sakaridis2021acdc}} \\ 
                 \cmidrule[0.8pt]{2-5}

               & DeepLabv3+ \cite{chen2017rethinking} & RefineNet-lw \cite{nekrasov2018light}& RefineNet-lw \cite{nekrasov2018light} &  RefineNet-lw \cite{nekrasov2018light}  \\ 
            \midrule
            \textit{LQ} & \textit{40.36} & \textit{40.75} & \textit{65.20} & \textit{28.30} \\
            \hdashline
            DIP \cite{liu2022image}                 & 57.17 & 57.67 & \underline{67.81} & \underline{38.19} \\
            DIP* \cite{liu2022image}                & 51.81 & 50.35 & 67.16 & 32.98 \\
            URIE \cite{son2020urie}                 & 55.88 & 51.45 & 65.93 & 37.90 \\
            URIE* \cite{son2020urie}                & 50.56 & 48.23 & 65.93 & 32.71 \\
            NAFNet \cite{chen2022simple}            & \underline{58.41} & \underline{58.19} & 66.06 & 37.59 \\
            NAFNet* \cite{chen2022simple}           & 51.91 & 53.29 & 65.40 & 36.03 \\
            PromptIR \cite{potlapalli2024promptir}  & 58.05 & 57.54 & 66.76 & 37.86 \\
            PromptIR* \cite{potlapalli2024promptir} & 54.67 & 52.25 & 63.44 & 35.51 \\
            DiffBIR \cite{lin2023diffbir}           & 52.49 & 53.68 & 66.29 & 36.28 \\
            DiffBIR* \cite{lin2023diffbir}          & 48.90 & 48.56 & 63.26 & 33.12 \\
            DiffUIR \cite{zheng2024selective}       & 51.28 & 51.46 & 66.24 & 35.78 \\
            DiffUIR* \cite{zheng2024selective}      & 47.92 & 45.01 & 62.82 & 34.83 \\
            \rowcolor{LightCyan} \textbf{\method{}} & \textbf{66.05} & \textbf{65.73} & \textbf{70.77} & \textbf{39.27} \\
            \hdashline
            \textit{HQ} & \textit{75.64} & \textit{75.66} & \textit{75.66} & - \\
            \bottomrule[1.3pt]
        \end{tabular}
    }
    \caption{\textbf{Performance comparison of existing methods on seen and unseen datasets for semantic segmentation.} Results are reported in terms of mIoU.}
    \label{tab:seg_cityscapes}
\end{table}

\begin{figure}[t!]
  \centering
  \setlength{\tabcolsep}{1pt}
  \footnotesize
    \begin{tabular}{cccccc}
        \includegraphics[width=.09\textwidth, height=1.4cm]{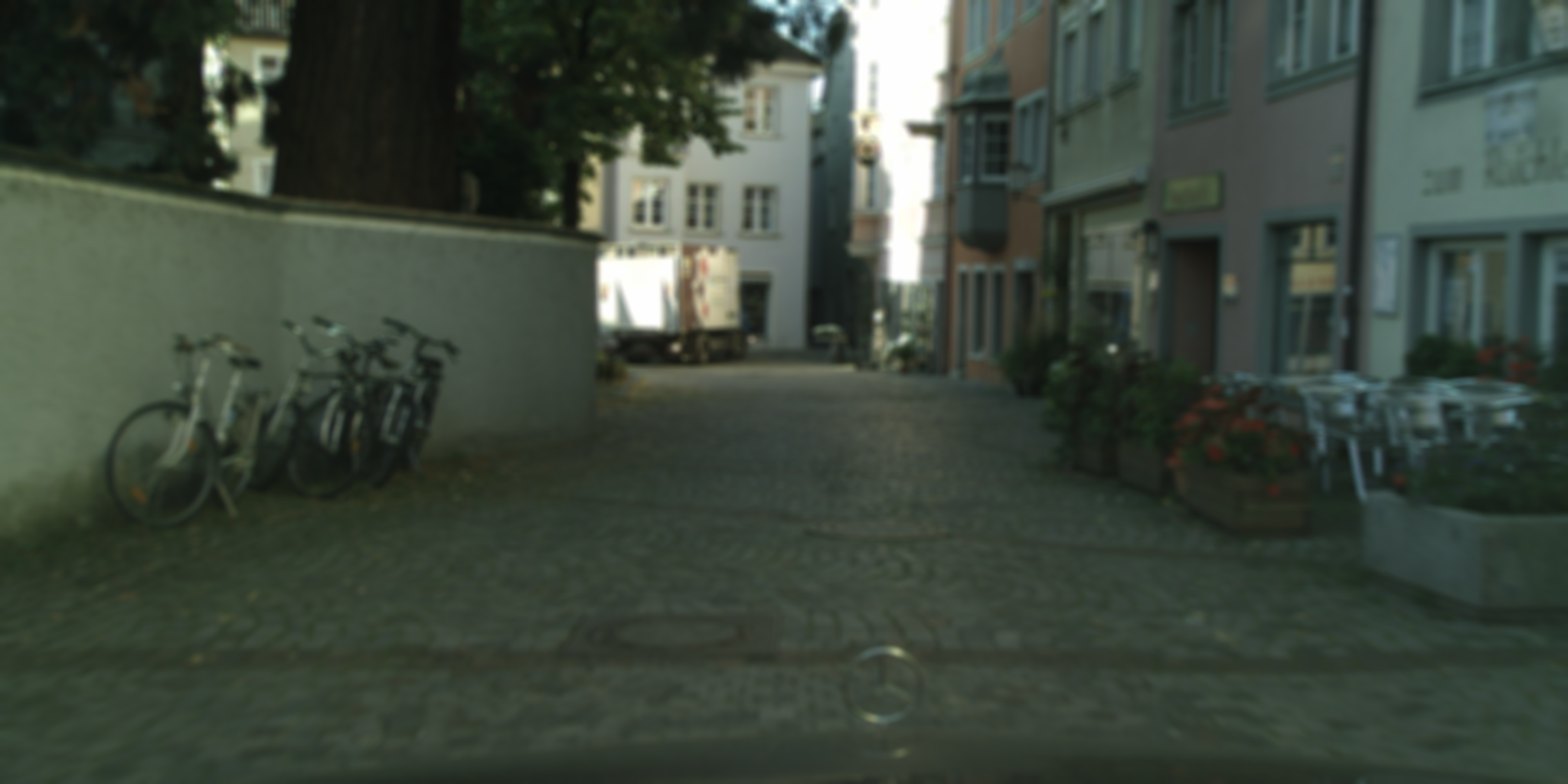} &
        \includegraphics[width=.09\textwidth, height=1.4cm]{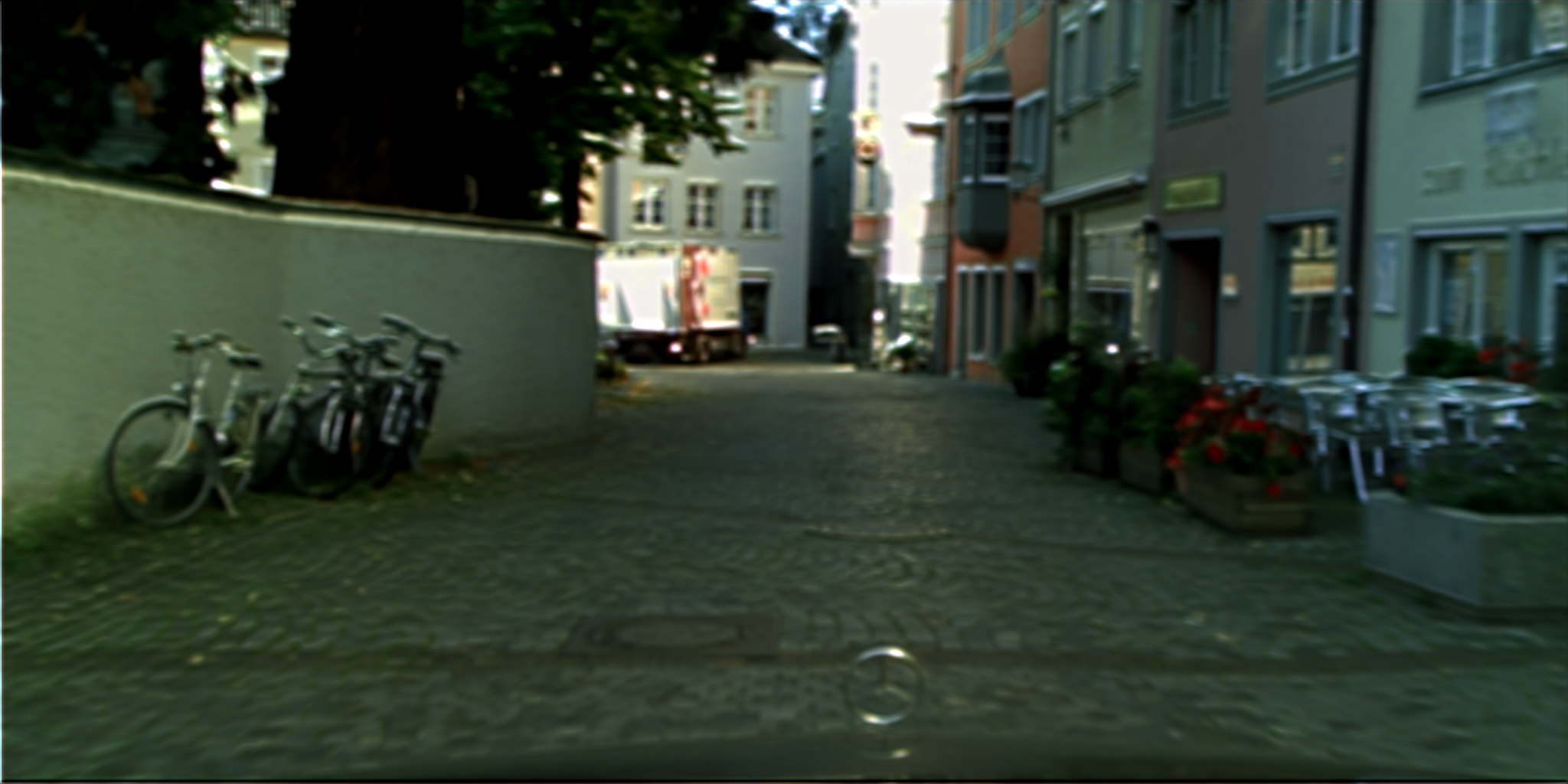} &
        \includegraphics[width=.09\textwidth, height=1.4cm]{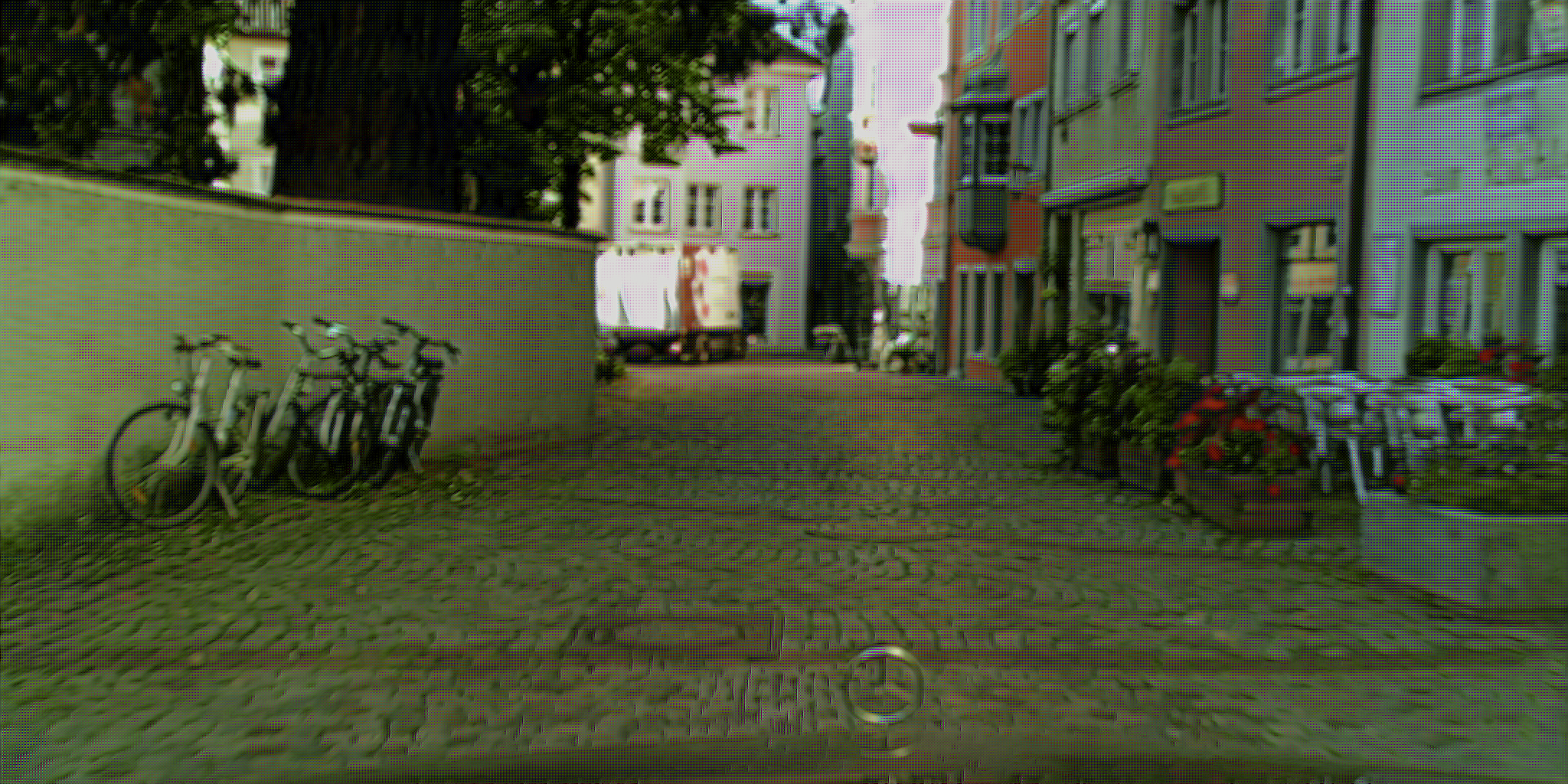} &
        \includegraphics[width=.09\textwidth, height=1.4cm]{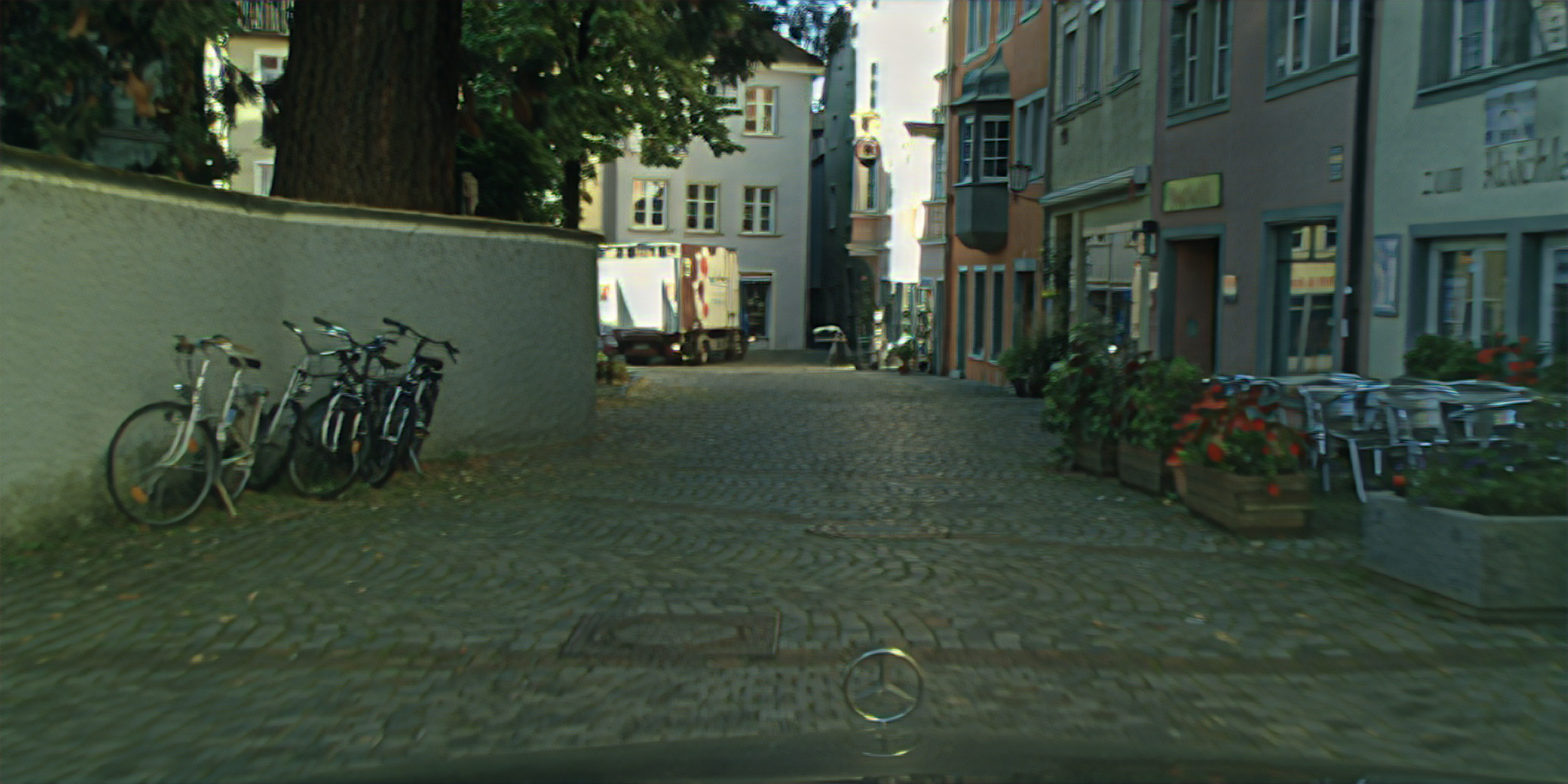} &
        \includegraphics[width=.09\textwidth, height=1.4cm]{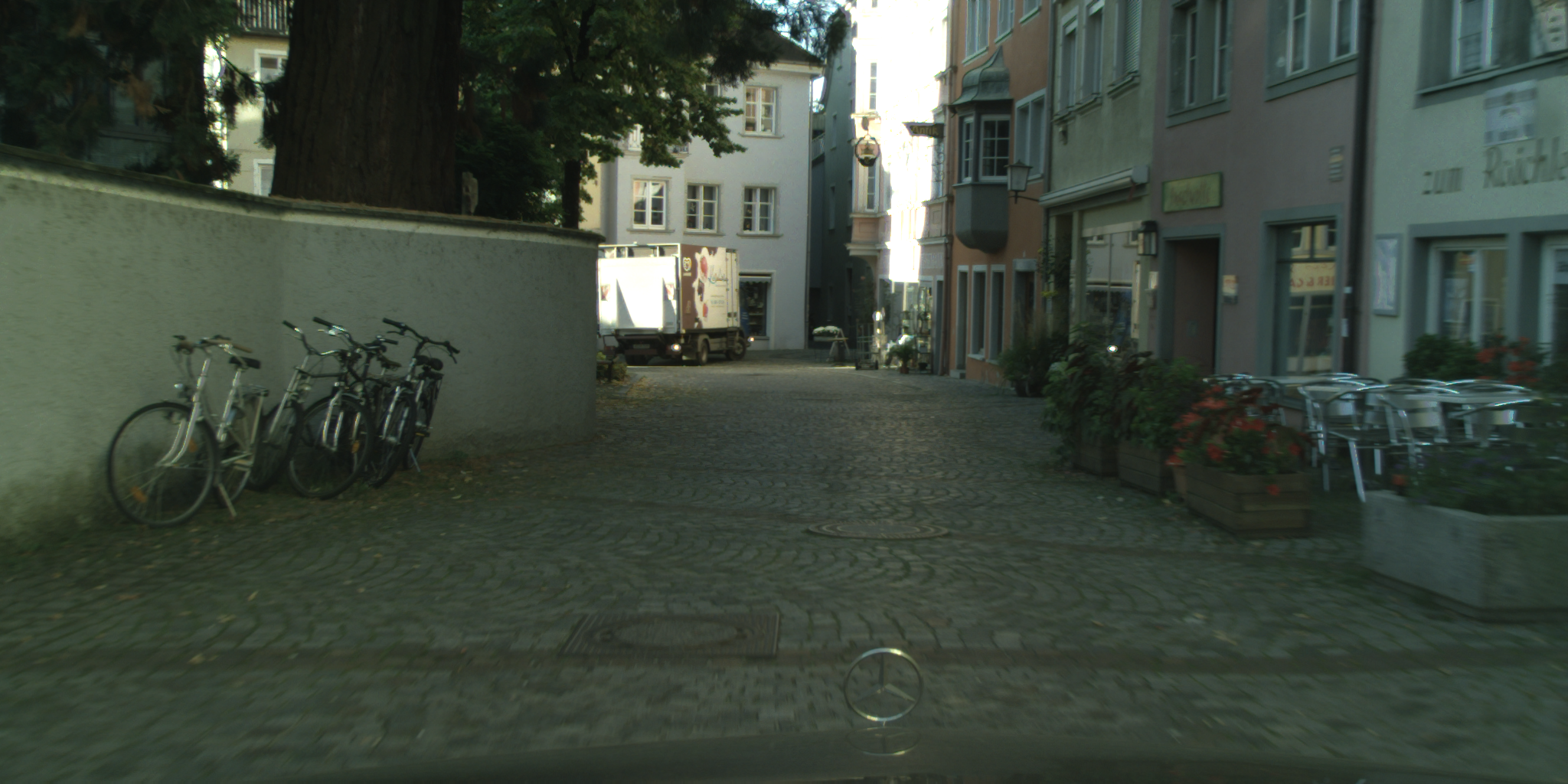} \\

        \includegraphics[width=.09\textwidth, height=1.4cm]{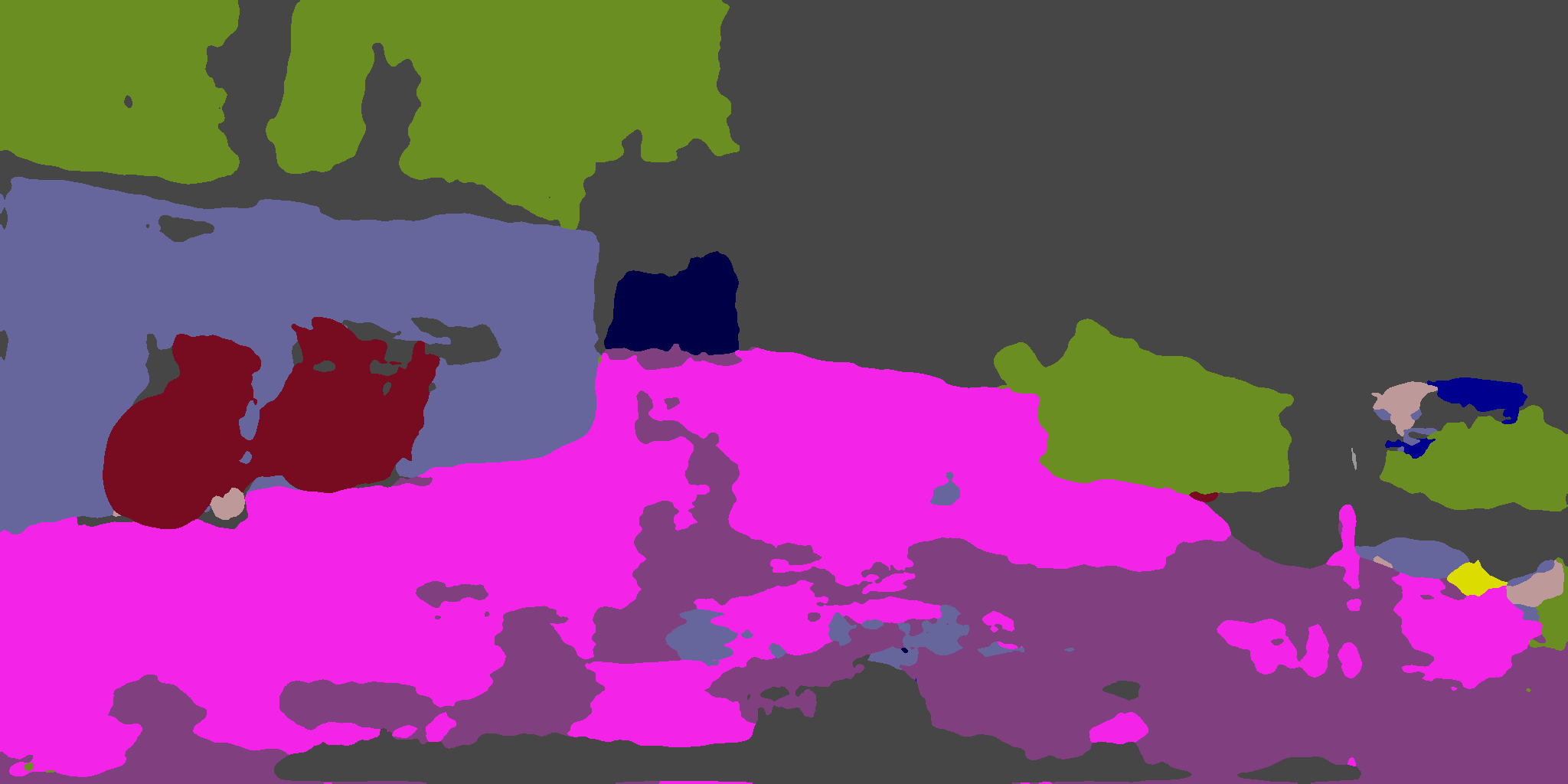} &
        \includegraphics[width=.09\textwidth, height=1.4cm]{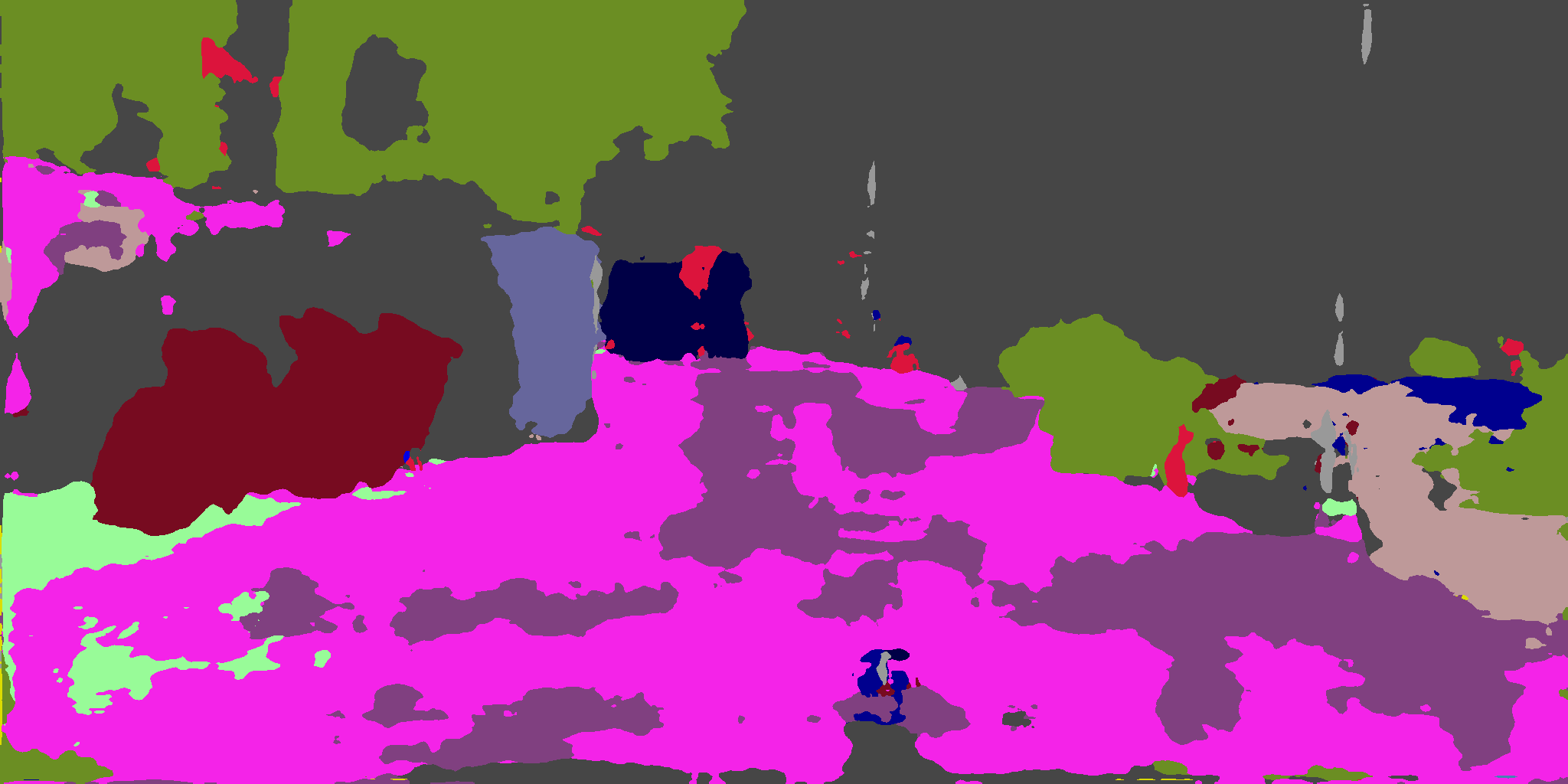} &
        \includegraphics[width=.09\textwidth, height=1.4cm]{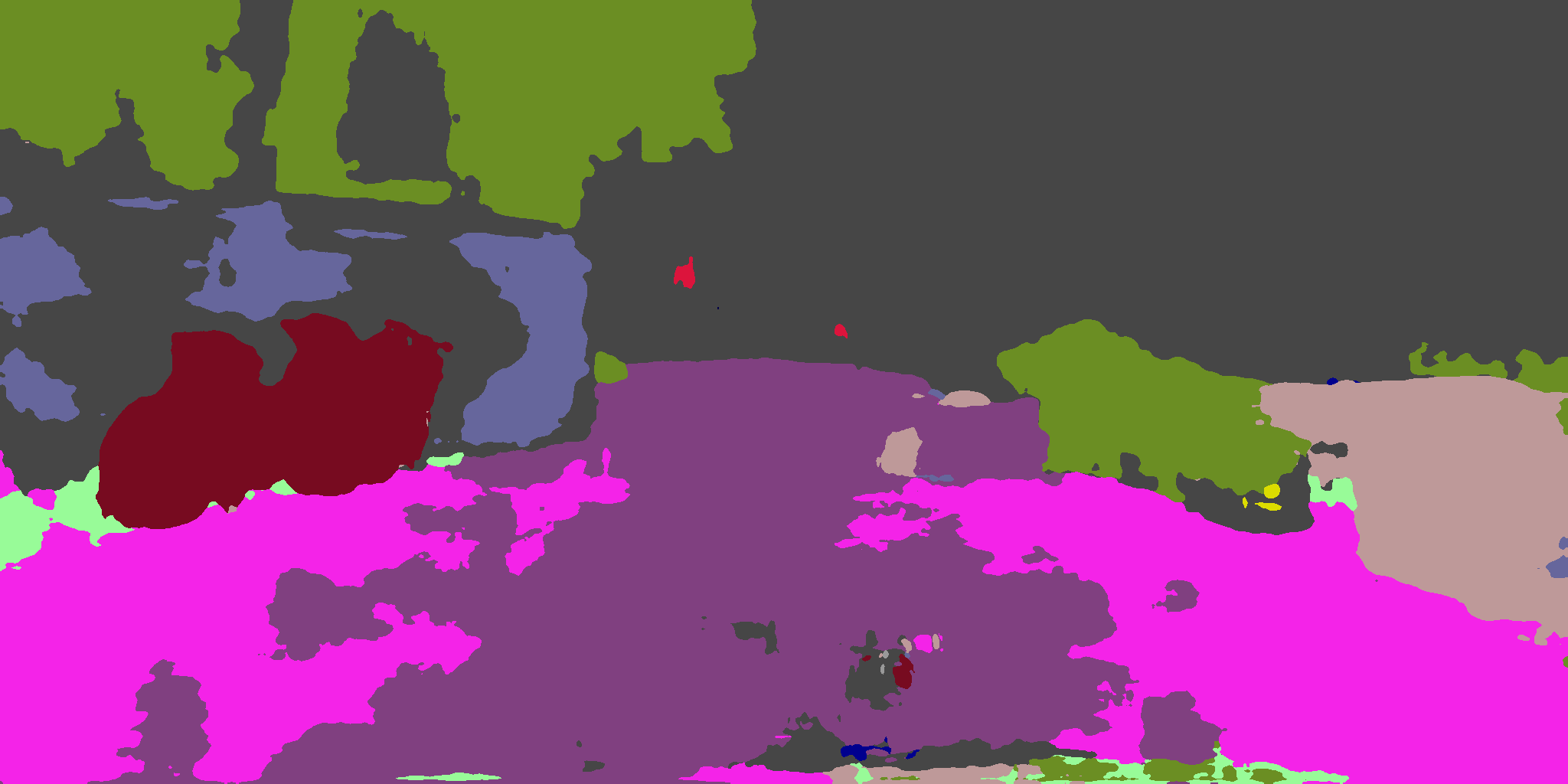} &
        \includegraphics[width=.09\textwidth, height=1.4cm]{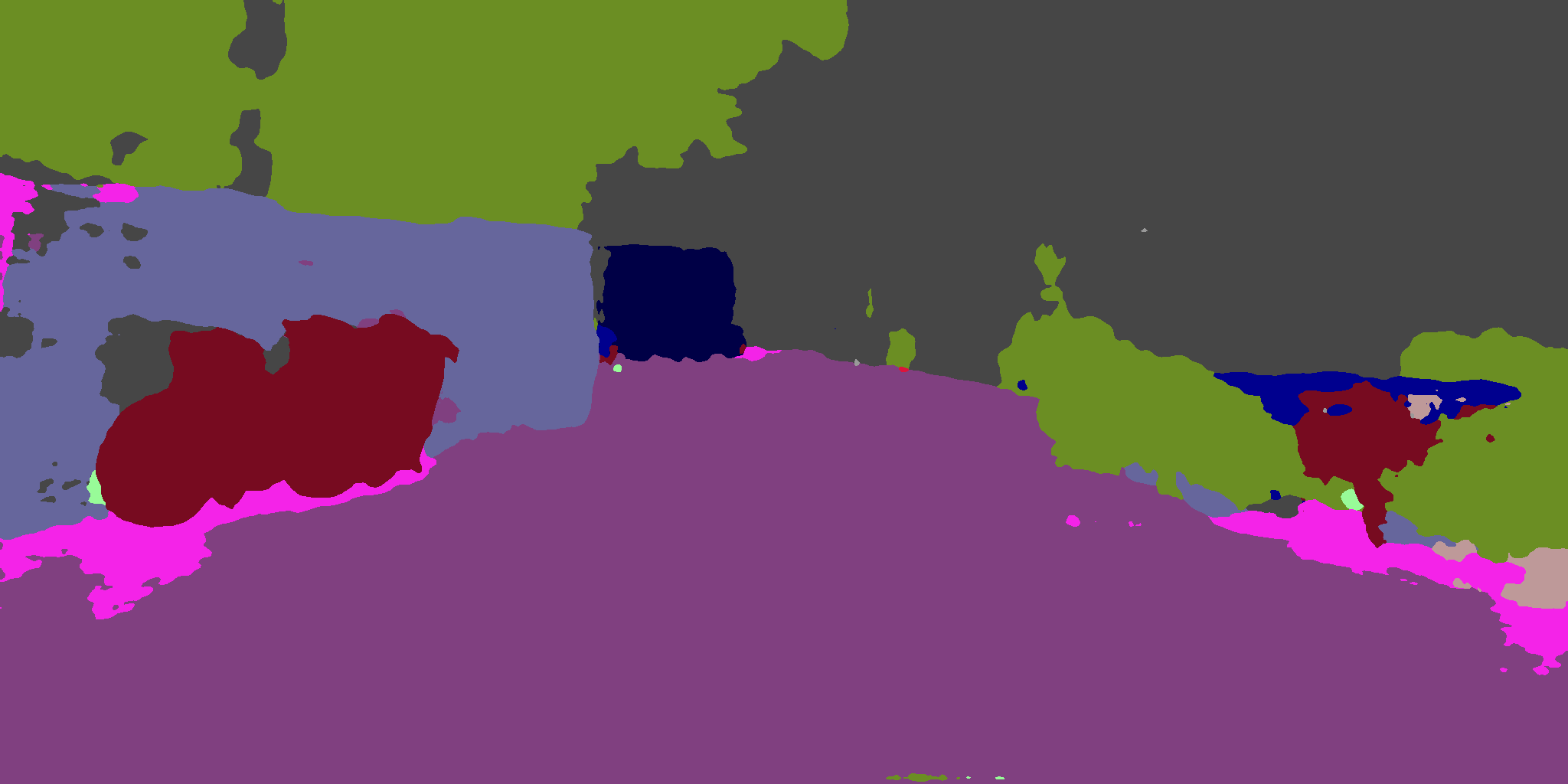} &
        \includegraphics[width=.09\textwidth, height=1.4cm]{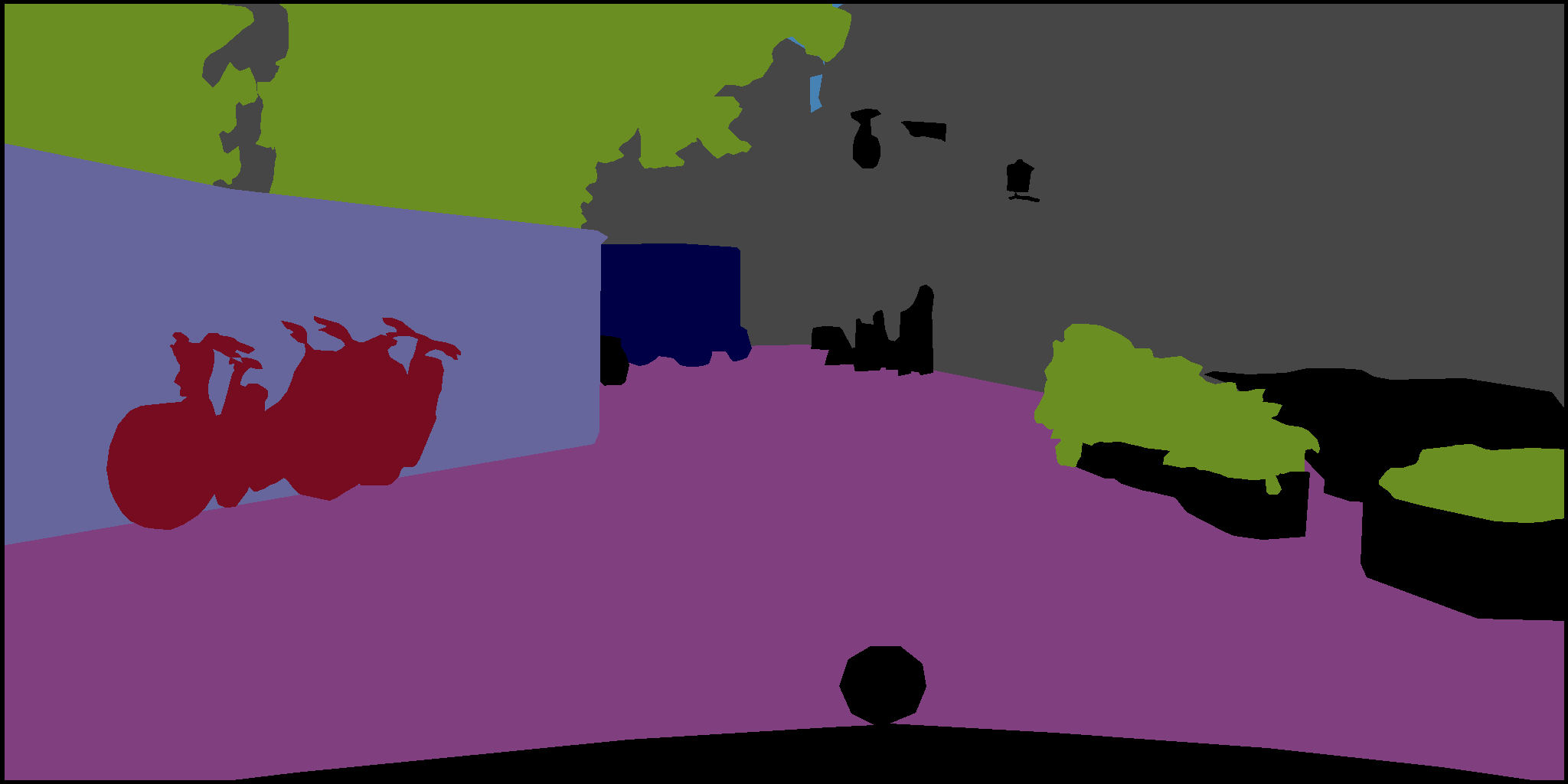} \\

        \includegraphics[width=.09\textwidth, height=1.4cm]{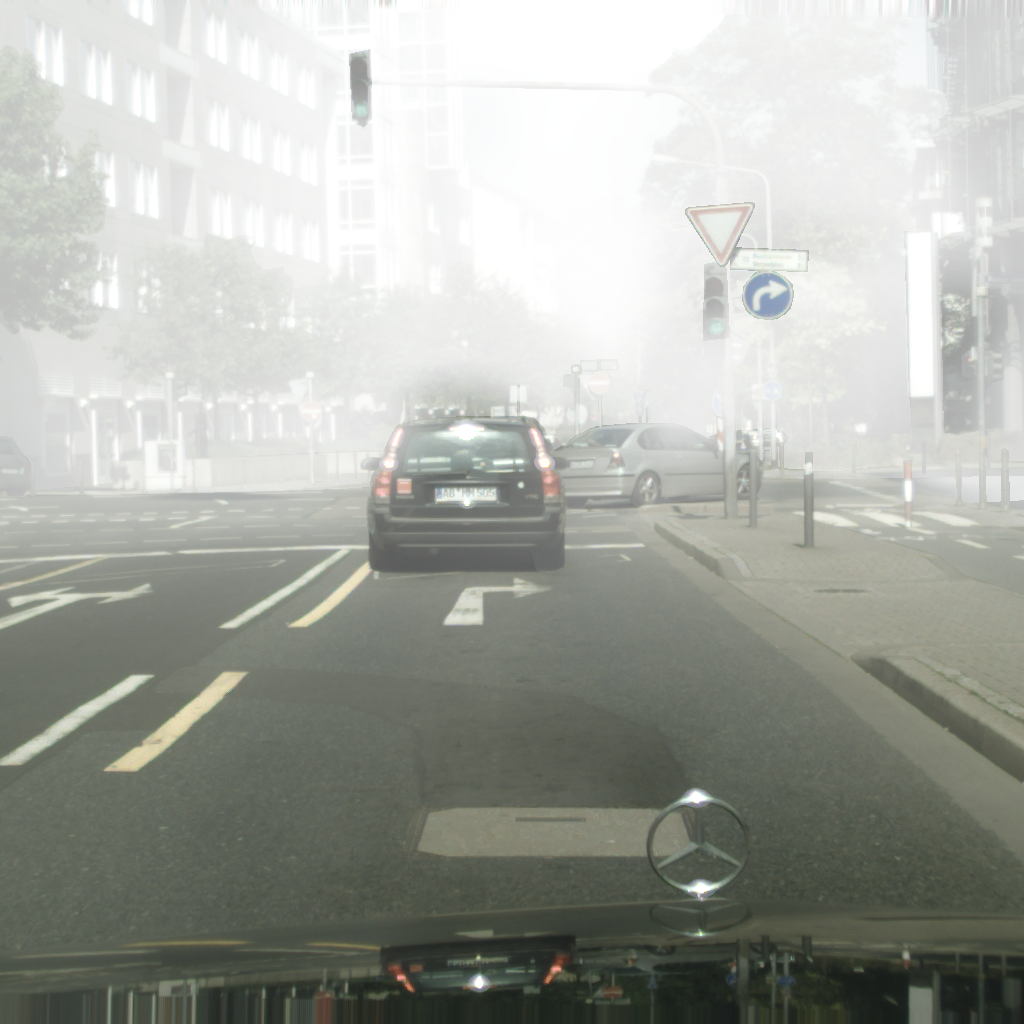} &
        \includegraphics[width=.09\textwidth, height=1.4cm]{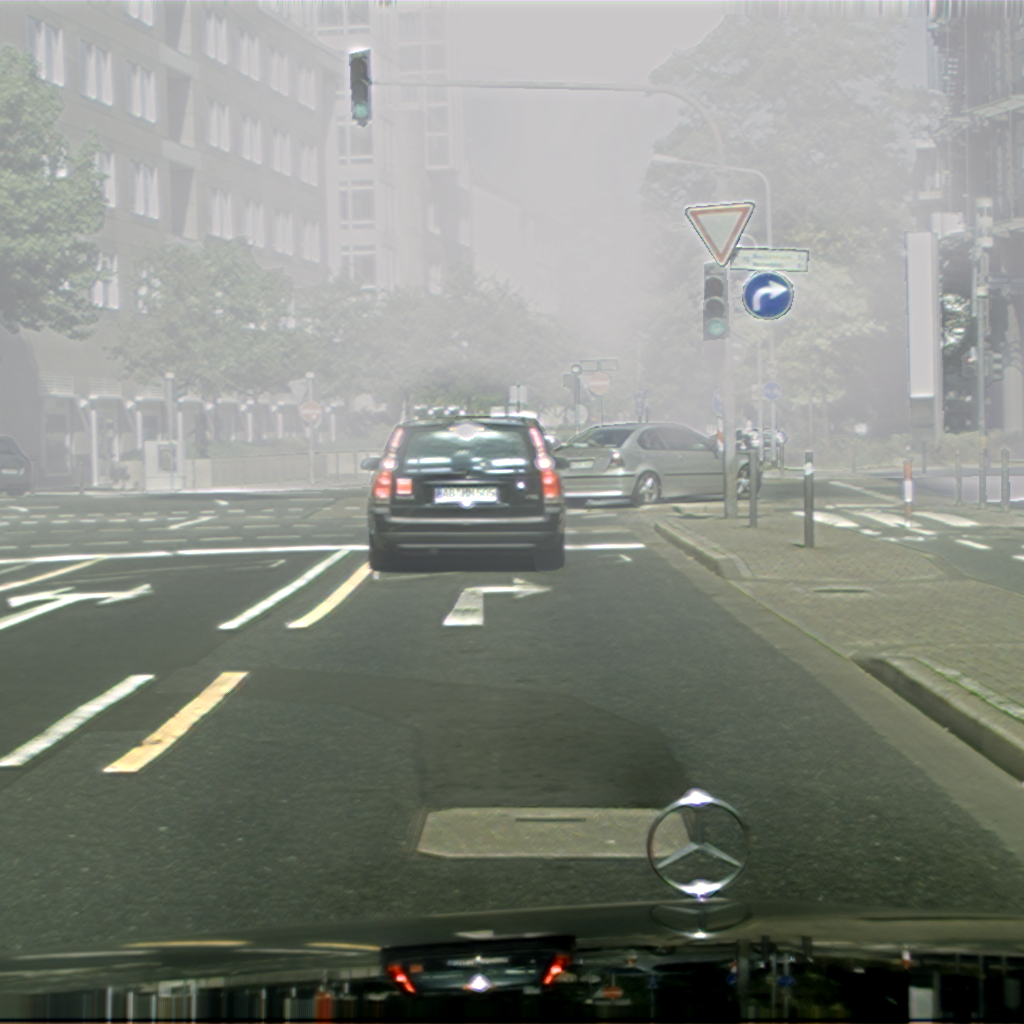} &
        \includegraphics[width=.09\textwidth, height=1.4cm]{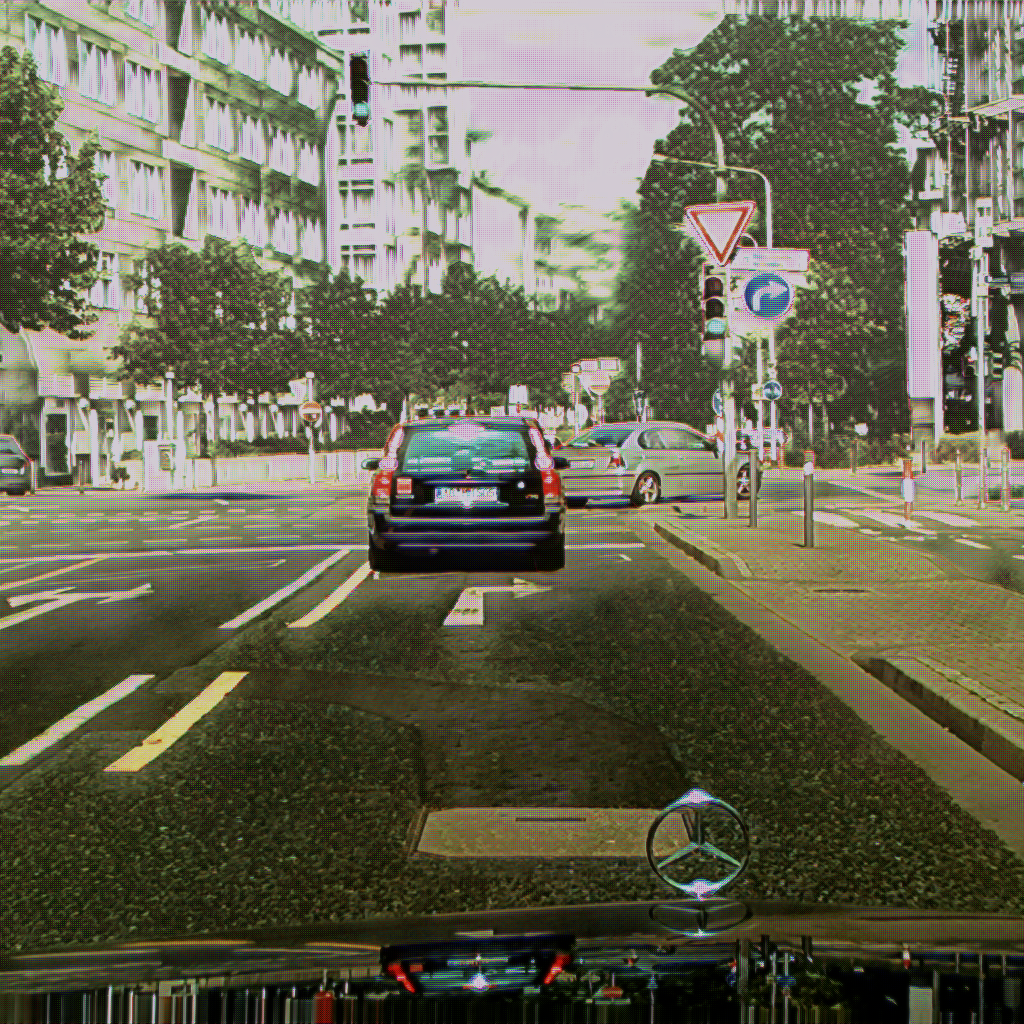} &
        \includegraphics[width=.09\textwidth, height=1.4cm]{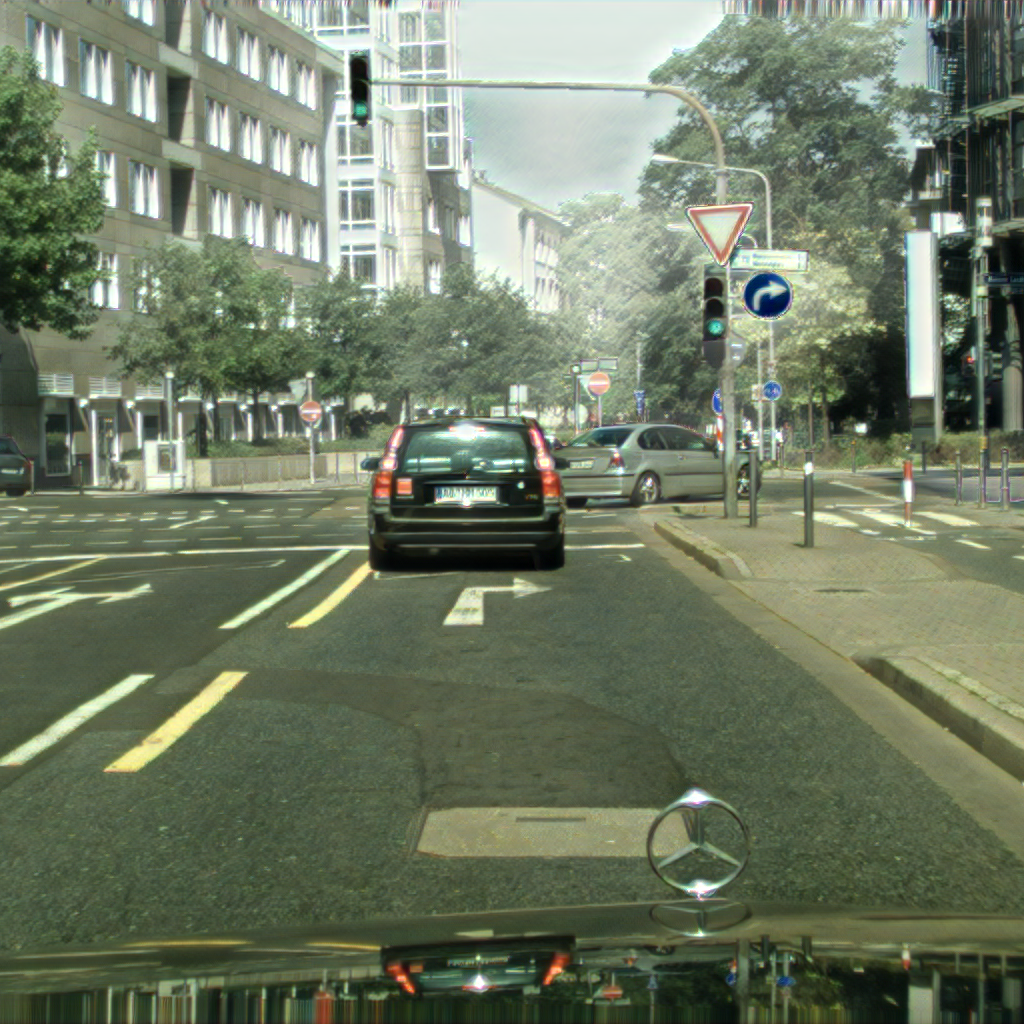} &
        \includegraphics[width=.09\textwidth, height=1.4cm]{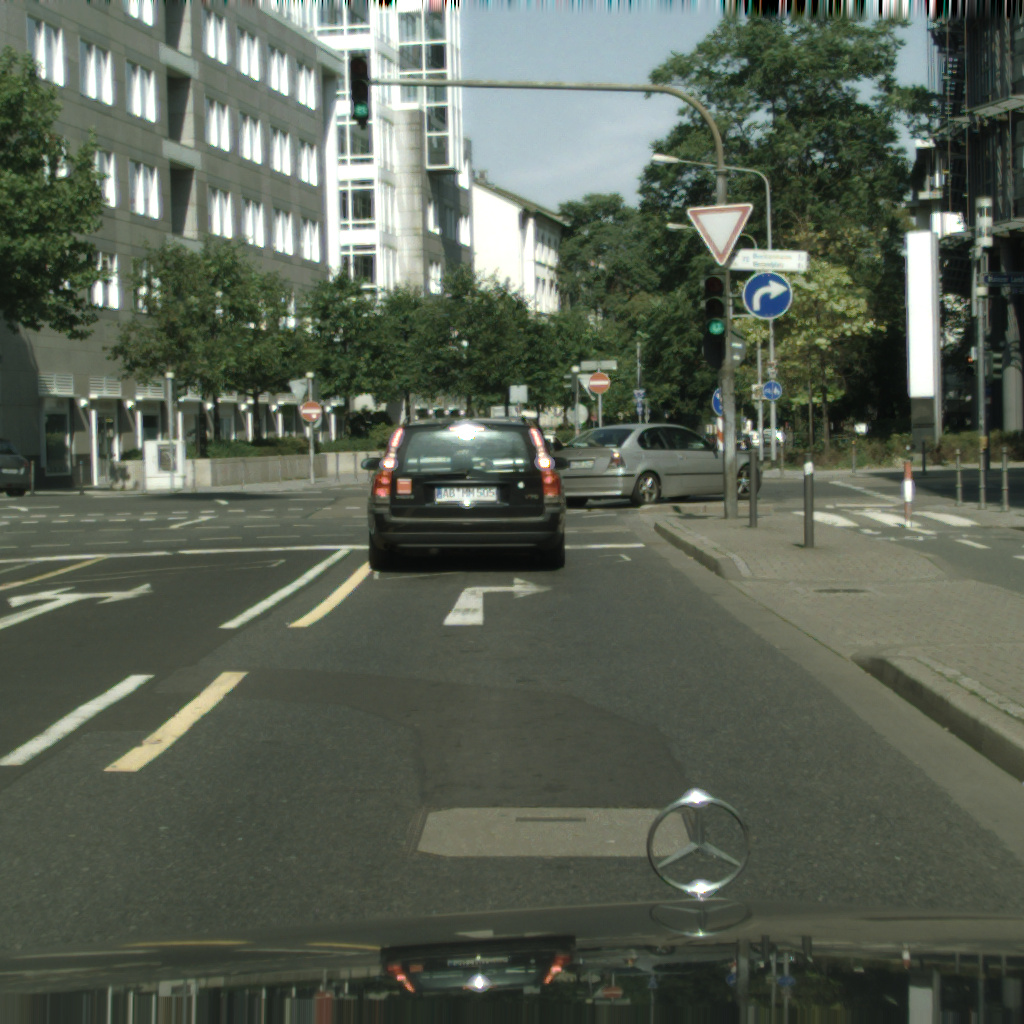} \\

        \includegraphics[width=.09\textwidth, height=1.4cm]{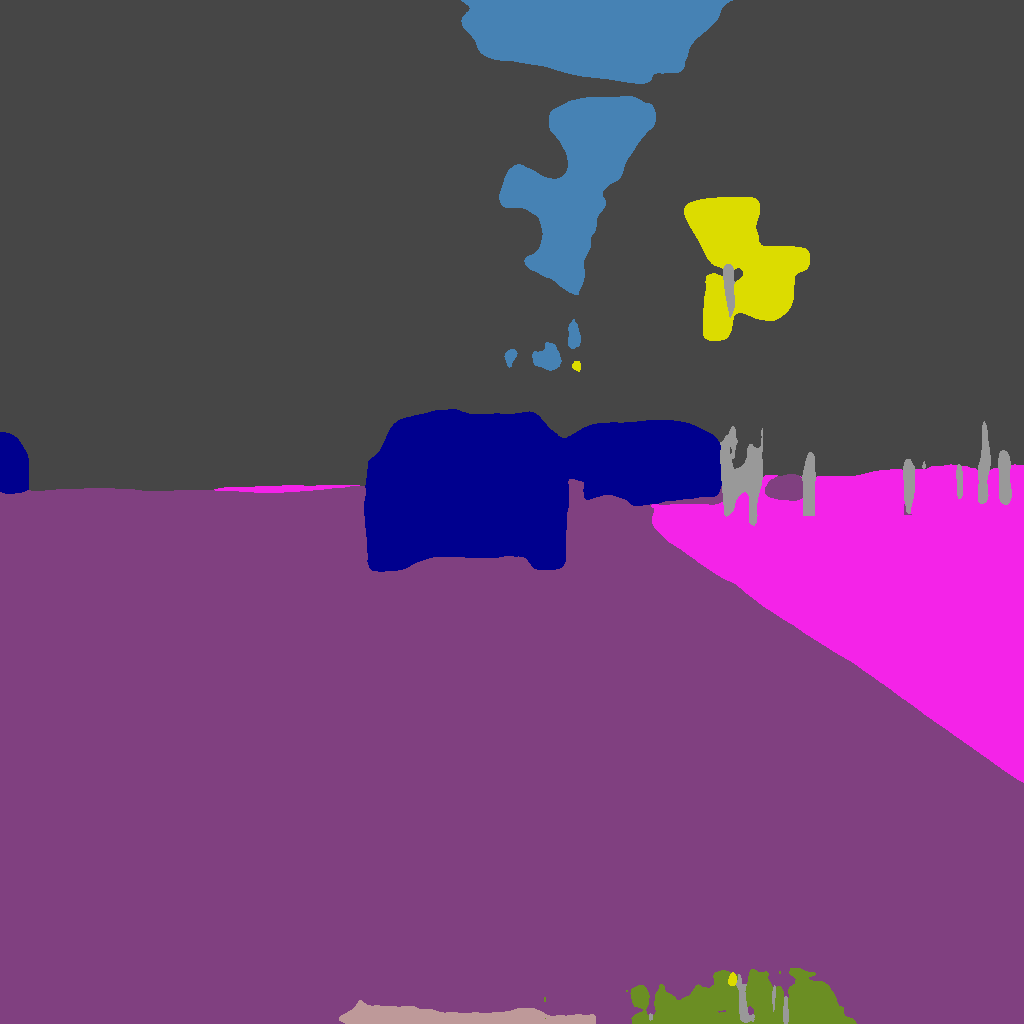} &
        \includegraphics[width=.09\textwidth, height=1.4cm]{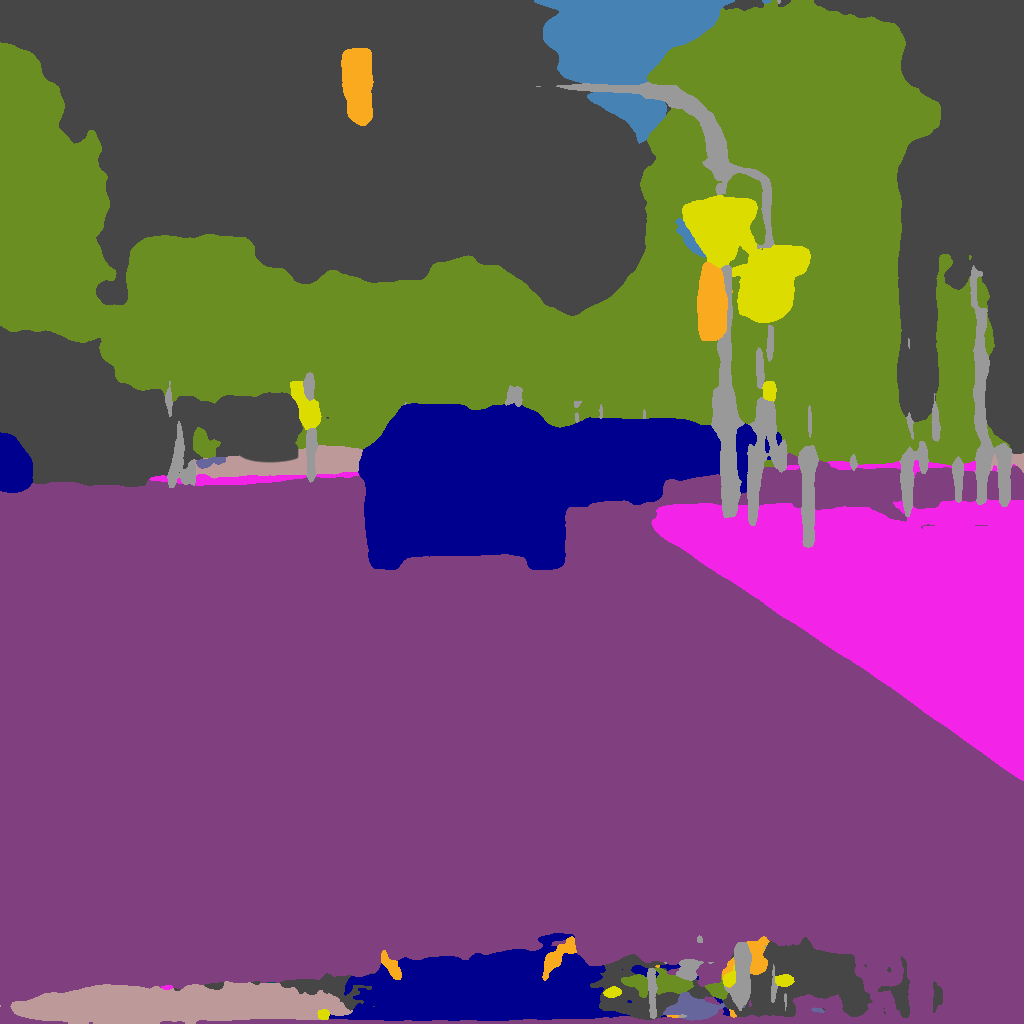} &
        \includegraphics[width=.09\textwidth, height=1.4cm]{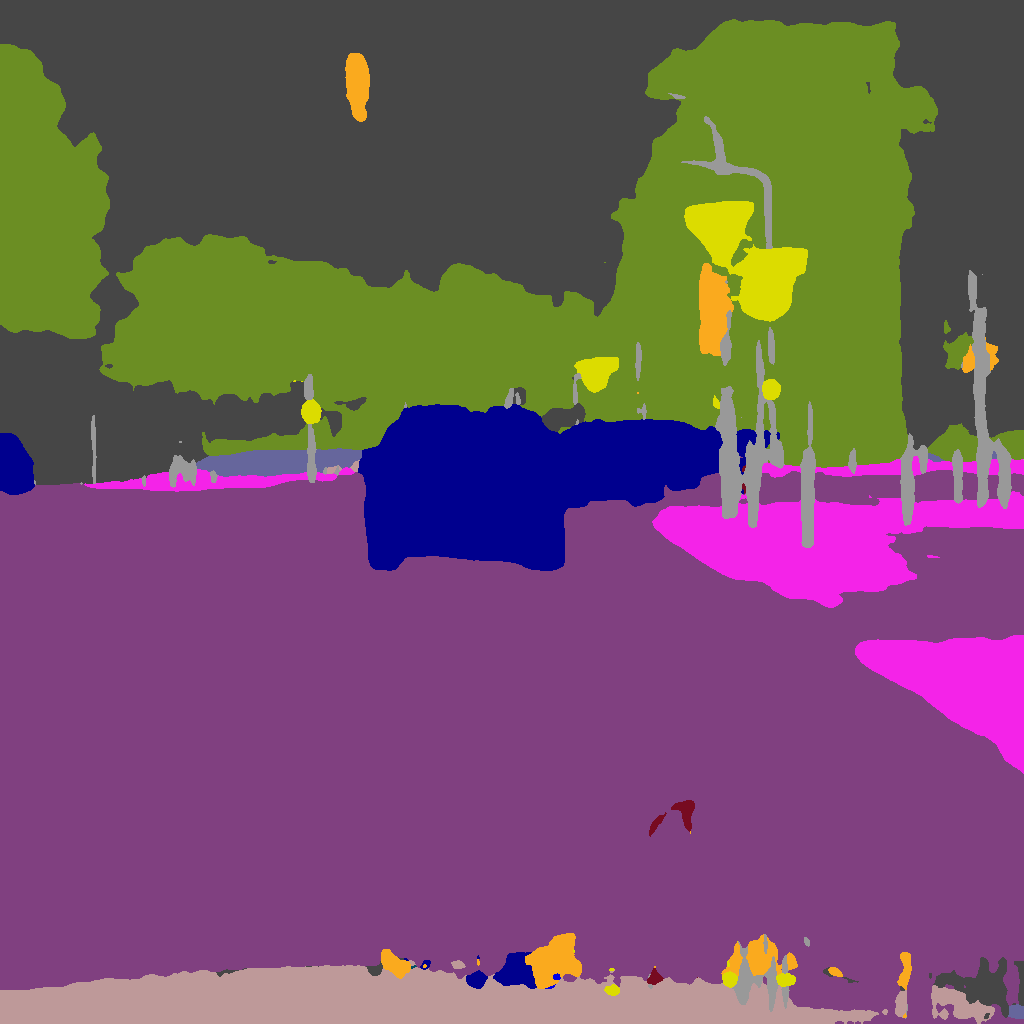} &
        \includegraphics[width=.09\textwidth, height=1.4cm]{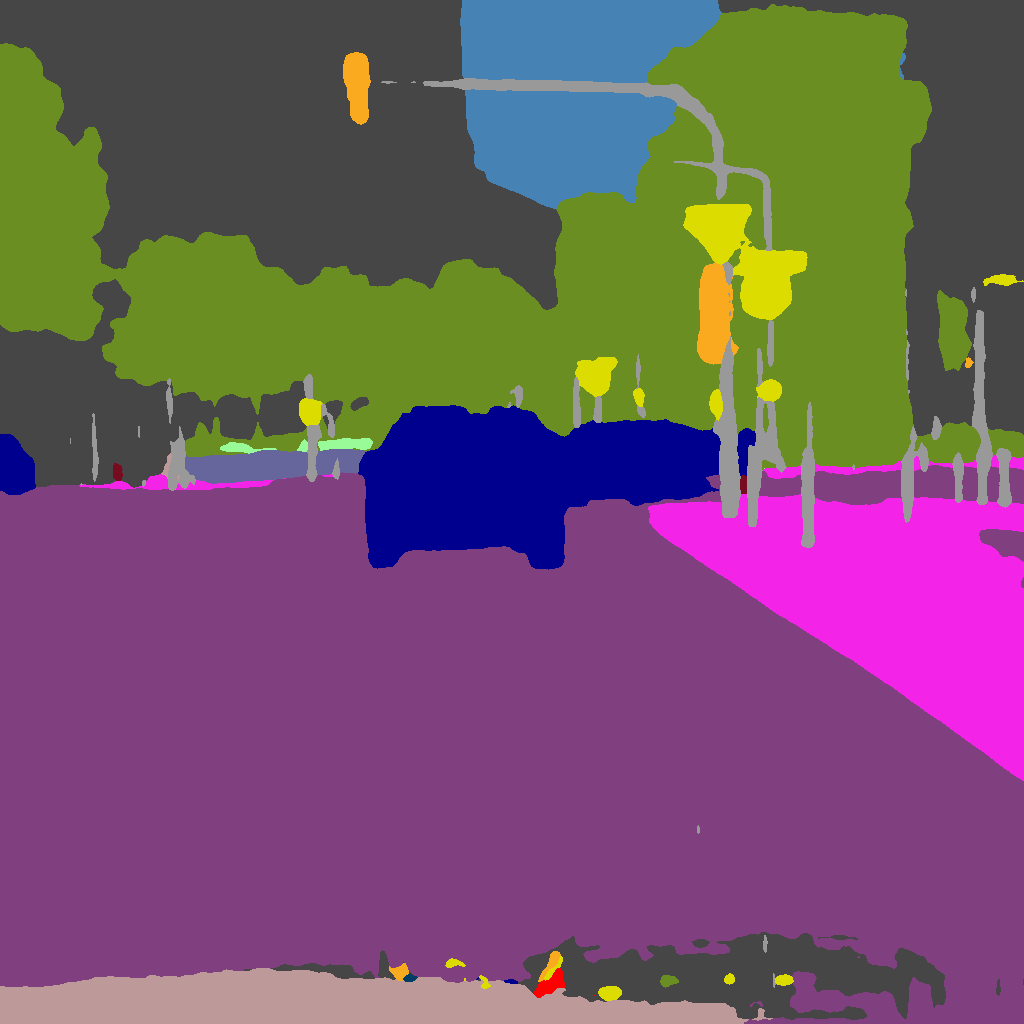} &
        \includegraphics[width=.09\textwidth, height=1.4cm]{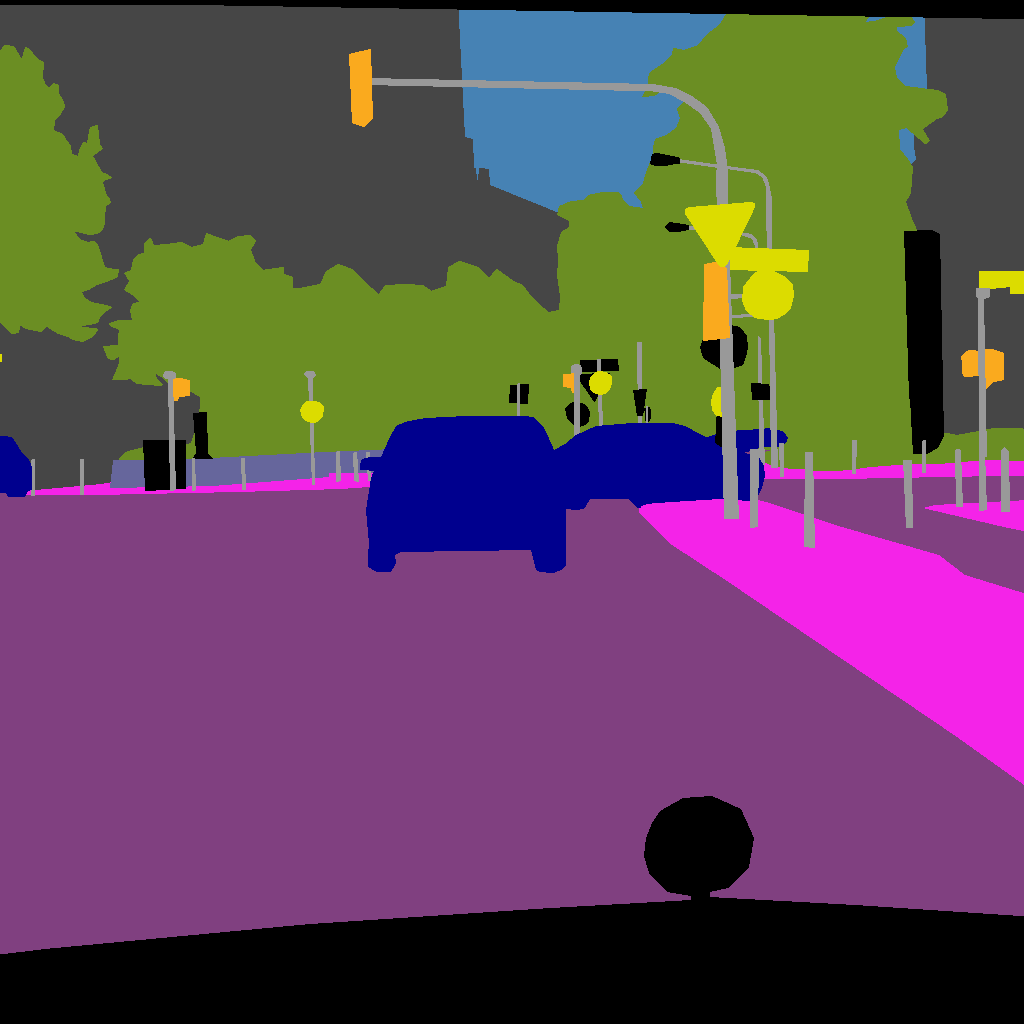} \\
        LQ & URIE & PromptIR & UniRestore & HQ  \\
    \end{tabular}
    \caption{\textbf{Qualitative analysis of semantic segmentation.} The first and third rows present the input images, while the second and fourth rows display the corresponding segmentation results.}
    \label{fig:seg_vis}
\end{figure}

\subsection{Task-oriented Image Restoration}
\noindent\textbf{Image Classification.} 
During training, \method{} employs ResNet-50~\cite{he2016deep} as the recognition model. For evaluation, both ResNet-50~\cite{he2016deep} and ViT-B~\cite{dosovitskiy2020image} serve as recognition backbones on the restored images. All recognition models are pre-trained on the training sets of their corresponding classification datasets without the degradation synthesis process. 

\tabref{tab:cls_performance} demonstrates \method{}'s effectiveness in enhancing image classification performance compared to existing image restoration models, achieving accuracy comparable to that obtained using high-quality ground truth inputs (\textit{i.e.}, HQ). Moreover, PIR methods show limited performance in classification tasks because they are optimized for human perception, which does not guarantee recognition accuracy. However, when these methods are trained on multiple downstream tasks, classification performance decreases. This decline may be due to these methods' potential inability to effectively handle multiple downstream tasks simultaneously. As detailed in \tabref{tab:cls_performance}, on the CUB dataset~\cite{wah2011caltech}, \method{} enhances classification accuracy by 20.01\% for ResNet-50~\cite{he2016deep} and by 15.96\% for ViT-B~\cite{dosovitskiy2020image} in scenarios involving unseen images. \figref{fig:cls_vis} visually demonstrates that when images restored by \method{} are used as inputs for recognition, their activation maps align more closely with those from high-quality ground truth images used as inputs for recognition.

\noindent\textbf{Semantic Segmentation.} 
In the training stage, we adopt DeepLabv3+ \cite{chen2017rethinking} as the segmentation model, while for evaluation, we employ both DeepLabv3+ \cite{chen2017rethinking} and RefineNet-lw \cite{nekrasov2018light}. Both models are pretrained on the Cityscapes~\cite{cordts2016cityscapes} training set. As shown in \tabref{tab:seg_cityscapes}, \method{} achieves decent performance in semantic segmentation on both seen datasets and the unseen dataset. These results underscore \method{}'s ability to restore fine-grained details crucial for semantic segmentation tasks, demonstrating that the TFA modules integrate diffusion features with restored features to produce restored images with high-quality feature representations. Similar to the results observed in classification, PIR methods show limited performance in semantic segmentation compared to TIR methods, with performance decreasing in multiple downstream task scenarios. A qualitative comparison in \figref{fig:seg_vis} shows that images restored by \method{} enable segmentation models to generate more accurate object boundaries in the segmented results.

\subsection{Ablation Study}
To verify the effectiveness of the proposed modules, we conduct an ablation study. All experiments are evaluated on the DIV2K \cite{agustsson2017ntire} test set for PIR, the ImageNet \cite{deng2009imagenet} test set for image classification, and the Cityscapes \cite{sakaridis2018semantic} test set for semantic segmentation, with degradation synthesis applied in the above tasks. These three sets are the same as we used in the evaluation dataset.

\noindent\textbf{Effectiveness of Proposed Modules.}
We have established several configurations for our experiments: (i) Baseline: training the controller with a pre-trained Stable Diffusion model; (ii) \method{} w/o CFRM: using the vanilla encoder features without any restoration; (iii) \method{} w/o TFA: employing only the latent features from the denoising U-Net without adapting the encoder features; (iv) \method{}: incorporating all modules. \tabref{tab:ablation_\method{}} demonstrates that both CFRM and TFA significantly enhance performance across PIR and TIR scenarios.

\begin{table}[t!]
    \centering
    \scalebox{0.7}{
        \begin{tabular}{l|ccc} 
            \toprule[1.3pt]
            \multirow{2}{*}{Methods} & PIR & Cls & Seg \\ 
              & PSNR $\uparrow$ & ACC $\uparrow$ &  mIoU $\uparrow$ \\ 
            \midrule
             Baseline            & 19.35 & 57.65 & 46.76 \\
             \method{} w/o CFRM  & 21.43 & 63.10 & 55.48 \\
             \method{} w/o TFA   & 22.16 & 64.25 & 58.13 \\
             \rowcolor{LightCyan} \textbf{\method{}}  & \textbf{24.32} & \textbf{71.65} & \textbf{66.05} \\
            \bottomrule[1.3pt]
        \end{tabular}
    }
    \caption{\textbf{Effectiveness of CFRM and TFA in \method{}.}}
    \label{tab:ablation_\method{}}
\end{table}

\begin{table}[t!]
    \centering
    \scalebox{0.70}{
        \begin{tabular}{l|c|ccc}
            \toprule[1.3pt]
              \multirow{2}{*}{Methods}           & \multirow{2}{*}{\# of Tuned Parameters} & PIR             & Cls            & Seg \\ 
              &      & PSNR $\uparrow$ & ACC $\uparrow$ &  mIoU $\uparrow$ \\ 
            \midrule
             Multiple Adapters     & 65.17M & 23.06 & 68.95 & 64.64 \\
             Multiple TFAs         & 63.03M & \textbf{25.48} & 71.20 & 65.78 \\
            \method{}-SP           & 21.01M & 23.91 & 70.05 & 64.99 \\
             \rowcolor{LightCyan}  \textbf{\method{}}  & 21.03M & 24.32 & \textbf{71.65} & \textbf{66.05} \\
            \bottomrule[1.3pt]
        \end{tabular}
    }
    \caption{\textbf{Comparative analysis of different TFA variants.}}
    \label{tab:ablation_TFA}
\end{table}

\begin{table}[t!]
    \centering
    \scalebox{0.70}{
        \begin{tabular}{c|cccc}
            \toprule[1.3pt]
            Method         & LQ         & DIP \cite{liu2022image}   &  PromptIR \cite{potlapalli2024promptir} &  \textbf{\method{}}   \\ 
            \midrule
            mAP $\uparrow$ & 45.63  & 54.29  &  50.61  & \textbf{58.06}      \\
            \bottomrule[1.3pt]
        \end{tabular}
    }
    \caption{\textbf{Performance of Extendability Tested on Object Detection Using the RTTS \cite{li2018benchmarking} Dataset.} }
    \label{tab:ablation_det}
\end{table}

\noindent\textbf{Investigation of TFA.}
To further investigate the effectiveness of the TFA module in multi-task scenarios, we conducted experiments with four variants: (i) Multiple Adapters: concatenates the output of the denoising U-Net with the restored features from CFRM and processes them through the same number of convolutional blocks as in TFA; (ii) Multiple TFAs: optimizes each task with its own TFA; (iii) \method{}-SP: employs a single TFA with a single prompt for all tasks; (iv) \method{}: utilizes one TFA with specific prompts for each task.

The results are shown in \tabref{tab:ablation_TFA}. The Multi-TFA outperforms Multi-Adapter, indicating the importance of dynamically fusing features by utilizing an updated prompt from the previous layer. Although \method{}-SP requires the fewest parameters to be tuned, its performance is inferior to that of \method{}, highlighting the significance of having a specific prompt for each task. \method{} delivers performance comparable to Multi-TFA in PIR tasks and better performance in TIR tasks. This may be attributed to the single TFA block's ability to update using different objectives simultaneously, unlike in Multi-TFA, where each TFA is updated with only one objective. This also suggests that knowledge from various TIR tasks can potentially benefit other TIR tasks. Additionally, the number of parameters needing updates does not significantly increase with the number of tasks, indicating that the proposed TFA structure effectively balances scalability and performance.





\noindent\textbf{Extendability Evaluation.}
To validate the extensibility of \method{}, we incorporate an additional downstream task—object detection based on the model trained for PIR, image classification, and semantic segmentation. Specifically, we use a RetinaNet \cite{lin2017refinenet} pre-trained on the COCO \cite{lin2014microsoft} as the backbone. We randomly select 69,242 images from the COCO training set and synthetic the degradation as our training set. As outlined in Section~\ref{sec:twostage}, we utilize the current model configuration and update only with a new learnable prompt, optimizing it using the object detection loss. We then evaluate the object detection performance of \method{} on the RTTS \cite{li2018benchmarking} dataset in comparison with other methods optimized concurrently for PIR, image classification, semantic segmentation, and object detection. As shown in~\tabref{tab:ablation_det}, \method{} achieves promising results in object detection. Additionally, compared to existing methods that require retraining models on complete task datasets, \method{} only needs fine-tuning of a prompt with new downstream data and optimizing with its specific objective. This highlights \method{}'s potential for extensibility to other downstream tasks using our designed TFA module.




%% file: Conclusion.tex
\section{Conclusion}
This paper introduces \method{}, an approach capable of addressing PIR and TIR simultaneously. Building on diffusion models, we propose adapting diffusion features for diverse applications. To achieve this, we introduce a complementary feature restoration module that restores features within the encoder and a task feature adapter that dynamically and efficiently combines these restored features with diffusion features for downstream tasks. Experimental results validate the effectiveness and extendability of \method{}, demonstrating its ability to alleviate the trade-offs associated with existing PIR and TIR methods.

\section*{Acknowledgements}
This research was supported by Taiwan's National Science and Technology Council under Grant NSTC 111-2221-E-002-136-MY3 and the Intelligence Advanced Research Projects Activity (IARPA) via Department of Interior/ Interior Business Center (DOI/IBC) contract number 140D0423C0074. The U.S. Government is authorized to reproduce and distribute reprints for Governmental purposes notwithstanding any copyright annotation thereon. Disclaimer: The views and conclusions contained herein are those of the authors and should not be interpreted as necessarily representing the official policies or endorsements, either expressed or implied, of IARPA, DOI/IBC, or the U.S. Government.